\documentclass[pdftex,twocolumn,epjc3]{svjour3}          % twocolumn
\pdfoutput=1
\RequirePackage[T1]{fontenc}
\smartqed  % flush right qed marks, e.g. at end of proof
\RequirePackage{graphicx}
\RequirePackage{mathptmx}      % use Times fonts if available on your TeX system
\RequirePackage{flushend}
\RequirePackage[numbers,sort&compress]{natbib}
\RequirePackage[colorlinks,citecolor=blue,urlcolor=blue,linkcolor=blue]{hyperref}
\usepackage{ifpdf}
\usepackage{amsmath,amssymb}
\usepackage{calc}
\usepackage{url}

\journalname{Eur. Phys. J. C}

\def \zc{z_{\textrm{cut}}}
\def \sd{\text{d}}
\def\la{\left\langle}
\def\ra{\right\rangle}

\newcommand{\ie}{i.e.\ }

\newcommand{\beq}{\begin{equation}}
\newcommand{\eeq}{\end{equation}}

\newcommand{\bdm}{\begin{displaymath}}
\newcommand{\edm}{\end{displaymath}}

\newcommand\f[2]{\frac{#1}{#2}}
\def\Ph{{\hat P}}

\def\beeq{\begin{eqnarray}}
\def\eeeq{\end{eqnarray}}
\def\ep{\epsilon}
\def\nn{\nonumber}
\def\lra{\leftrightarrow}

\newcommand{\bea}{\begin{eqnarray}}
\newcommand{\eea}{\end{eqnarray}}
\def\d{\partial}

\def \d{{\rm d} }
\def \d0 {D\O \;}

\def \zc{z_\text{cut}}
\def \event2{\textsc{Event2} }

\def\CMW{\textsc{cmw}}

\begin{document}

\title{Groomed jet mass as a direct probe of collinear parton dynamics
%\thanksref{t1}
}


\author{Daniele Anderle\thanksref{e1,addr1,addr2}
        \and
        Mrinal Dasgupta\thanksref{e2,addr3} %etc.
        \and
        Basem Kamal El-Menoufi\thanksref{e3,addr3}
        \and
        Marco Guzzi\thanksref{e4,addr4}
        \and
        Jack Helliwell\thanksref{e5,addr3}
}

%\thankstext[$\star$]{t1}{Thanks to the title}
\thankstext{e1}{e-mail: dpa@m.scnu.edu.cn}
\thankstext{e2}{e-mail: mrinal.dasgupta@manchester.ac.uk}
\thankstext{e3}{e-mail: basem.el-menoufi@manchester.ac.uk}
\thankstext{e4}{e-mail: mguzzi@kennesaw.edu}
\thankstext{e5}{e-mail: jack.helliwell@manchester.ac.uk}

\institute{Guangdong Provincial Key Laboratory of Nuclear Science, Institute of Quantum Matter, South China Normal University, Guangzhou 510006, China\label{addr1}
          \and
          Department of Physics and Astronomy, University of California, Los Angeles, CA 90095, U.S.A.\label{addr2}
          \and
          Consortium for Fundamental Physics, Department of Physics
            \& Astronomy, University of Manchester, Manchester M13 9PL, United Kingdom\label{addr3}
          \and
          Department of Physics, Kennesaw State University,
            Kennesaw, GA 30144, U.S.A.\label{addr4}
}

\date{Received: date / Accepted: date}

\maketitle
\keywords{QCD, Hadronic Colliders, Standard Model, Jets, Resummation}

\vspace{0.5 cm}

\begin{abstract}
We study the link between parton dynamics in the collinear limit and the logarithmically enhanced terms of the groomed jet mass distribution, for jets groomed with the modified mass-drop tagger (mMDT). While the leading-logarithmic (LL) result is linked to collinear evolution with leading-order splitting kernels, here we derive the NLL structure directly from triple-collinear splitting kernels. The  calculation we present is a fixed-order calculation in the triple-collinear limit, independent of resummation ingredients and methods. It therefore constitutes a powerful cross-check of the NLL results previously derived using the SCET formalism and provides much of the insight needed for resummation within the traditional QCD approach.
\end{abstract}

\section{Introduction}

The study of the substructure of jets is now an established and highly active component of LHC phenomenology. Following early work on the subject \cite{Seymour:1993mx,Butterworth:2002tt}, the true power of substructure methods for new particle discoveries in the
\\ boosted regime was first revealed over a decade ago \cite{Butterworth:2008iy}. This in turn led to an explosion of interest in the subject and rapid development of tools for tagging and grooming jets as well as their direct exploitation in experimental searches and studies at the LHC (see \cite{Abdesselam:2010pt,Altheimer:2012mn,Altheimer:2013yza,Asquith:2018igt} for reviews and further references).

 Jet substructure has also proved to be a fertile field for the development of calculations and concepts in QCD, aimed at improving the theoretical understanding of substructure. For a review of such work we refer the reader to Refs.~\cite{Larkoski:2017jix,Marzani:2019hun} and references therein.  An especially  important development in this context has been the invention of methods and observables that can be perturbatively calculated to high precision in a hadron collider environment and receive only modest non-perturbative corrections. This is a significant development for jet based studies which are generally subject to large non-perturbative uncertainties from hadronisation and the underlying event  (in addition to pile-up), as well as perturbative issues such as the presence of non-global logarithms where the resummation accuracy is still generally limited to leading-logarithms and the use of the leading $N_c$ limit \cite{Dasgupta:2001sh,Dasgupta:2002bw}.

A well-known set of observables, that eliminate some and ameliorate other issues obstructing precision for hadron collider jets, are those involving groomed jets where grooming is implemented through the modified mass-drop tagger (mMDT) \cite{Dasgupta:2013ihk,Dasgupta:2013via} or its subsequent  generalisation SoftDrop \cite{Larkoski:2014wba}. The use of a $\zc$ parameter as a threshold below which soft particles can be groomed away, eliminates non-global logarithms (NGLs) that otherwise appear in the ungroomed jet mass. Although NGLs still appear as logarithms in $\zc$, the phenomenologically relevant values of $\zc \sim 0.1$ imply that resummation of logarithms of $\zc$ is not strictly necessary. Moreover grooming via SoftDrop also considerably reduces the impact of non-perturbative effects relative to the \\
 ungroomed jet mass  \cite{Dasgupta:2013ihk}, virtually eliminating the underlying event  at high $p_T$ and leaving a modest hadronisation contribution amenable to analytic studies. These developments mean that theoretical calculations for jet observables with mMDT/SoftDrop
 grooming can be carried out to high precision and accurate comparisons can be made to LHC data, which is a program that has already been successfully established with several recent theoretical studies, experimental measurements and phenomenology \cite{Frye:2016aiz,Marzani:2017mva,Marzani:2017kqd,Baron:2018nfz,Bell:2018vaa, Kardos:2018kth, Kang:2018vgn, Marzani:2019evv,Kardos:2020gty,Kardos:2020ppl, Hoang:2019ceu,Aaboud_2018,Sirunyan_2018,Aad:2019vyi}.

 In this paper we focus on the mMDT (equivalently SoftDrop with  $\beta=0$) jet mass distribution in the region of small $\rho$, $\rho \ll \zc$, where $\rho$ is the normalised jet mass which for hadron collider jets is generally defined as  $\rho = \frac{m^2}{p_T^2 R^2}$, with $m^2$ the squared jet-mass, $p_T$ the jet's transverse momentum and $R$ the jet radius parameter. In addition to the removal of non-global logarithms and a considerable reduction in non-perturbative contributions mentioned above, the mMDT is unique in that the leading-logarithmic resummed result, for the jet mass  $\rho$, is single-logarithmic i.e. contains only terms $\alpha_s^n  \ln^n  \rho$ where the logarithms in $\rho$ are of pure {\emph{collinear origin}}. Such a structure is of course well-known for collinear evolution, both space-like and time-like, of partons via the DGLAP evolution equations for structure functions and fragmentation functions and for the small $R$ limit of jet cross-sections \cite{Dasgupta:2014yra} but is exceptional for observables like jet masses which in the ungroomed case receive double logarithmic terms at each perturbative order.

 The leading-logarithmic (LL) resummed result for the mMDT jet mass, amounting to a resummation of the single-logarithmic terms in $\rho$, was first derived in Ref.~\cite{Dasgupta:2013ihk}. The main result of that paper was written in the limit of small $\zc$ i.e. neglecting terms power suppressed in $\zc$. A result including finite $\zc$ corrections was also presented in the appendix of Ref.~\cite{Dasgupta:2013ihk} where complications arising from the leading parton flavour changing were accounted for through a matrix structure for the resulting resummation.
 An NLL result, in the small $\zc$ limit, was computed for the first time in Ref.~\cite{Frye:2016aiz} and most recently a resummation to NNLL accuracy in the same limit has been carried out \cite{Kardos:2020gty}.\footnote{Our description of the resummation accuracy for the  mMDT is consistent with the original reference  Ref.~\cite{Dasgupta:2013ihk} while more recent work uses a convention inspired by the general structure of SoftDrop jet mass, where single logarithms are referred to as NLL, in spite of the absence of double logarithmic terms. In this convention the work of Ref.~\cite{Frye:2016aiz} is NNLL while that of Ref.~\cite{Kardos:2020gty} achieves NNNLL accuracy.} Phenomenological  studies have also been carried out for \\ mMDT/SoftDrop jet masses where resummed calculations, supplemented by non-perturbative corrections, have been compared directly to experimental data. In the mMDT context the resummed calculations refer to LL  calculations including finite  $\zc$ effects and matching to full next-to--leading order (NLO) calculations \cite{Marzani:2017mva,Marzani:2017kqd} as well as calculations with NLL resummation but without finite $\zc$ corrections and matched to LO calculations \cite{Frye:2016aiz}. Generally good agreement with data has been observed in each case and we refer the reader to Refs.~\cite{Aaboud_2018,Sirunyan_2018,Aad:2019vyi} for further details.

 In the present article we take a fresh look at the NLL structure of the mMDT jet mass where previous work on the subject has been entirely within the context of Soft-Collinear Effective Theory (SCET) calculations \cite{Frye:2016aiz,Bell:2018vaa, Kang:2018vgn,Kardos:2020gty,Kardos:2020ppl}.  The original work of Ref.~\cite{Frye:2016aiz} recycled SCET calculations for soft functions with an energy veto \cite{Chien_2016,von_Manteuffel_2014} and calculated a correction for C/A jet clustering \cite{Dokshitzer_1997,wobisch1999hadronization} to derive the result for the mMDT.  In the current paper we shall approach this question from the viewpoint of QCD calculations by directly addressing the collinear nature of the $\ln \rho$ terms in the jet mass distribution. Thus while a leading-logarithmic resummed result can be derived in a strongly ordered picture  with collinear emissions widely separated in angle, the NLL structure with terms $\alpha_s^n \ln^{n-1} \rho$ stems from lifting the strong ordering on a pair of emissions. At order $\alpha_s^2$ this leads us to consider two emissions (i.e. three partons) within a jet, which are at small angles  $\theta^2 \lesssim \frac{\rho}{\zc} \ll 1$ but with no relative strong ordering between them. Such configurations are related to an NLL DGLAP evolution picture and the QCD matrix element can be approximated by triple-collinear splitting functions \cite{Campbell:1997hg,Catani:1998nv,Catani:1999ss}.  In the strongly-ordered limit the triple-collinear splitting functions reduce to a product of LO splitting functions thereby restoring the LL picture.

 There has been much recent interest and work towards the incorporation of higher-order splitting functions, including the triple-collinear and double-soft limits, in the context of parton shower algorithms \cite{Li:2016yez,Hoche:2017iem,Hoche:2017hno}. With the emergence of new research clarifying and pushing forward the logarithmic accuracy of parton showers \cite{Dasgupta:2018nvj,Dasgupta:2020fwr}, it is also of interest to understand in more detail the connection between higher-order splitting functions and the logarithmically enhanced terms for QCD observables.

 With the above points in mind, in this paper we compute the $\mathcal{O} \left(\alpha_s^2 \ln \rho\right)$ NLL term, for quark initiated jets, using the triple-collinear splitting functions and phase-space. Our calculation is a pure fixed-order computation, with the main approximation being the use of the collinear limit. In particular we do not make use of known ingredients from any resummation approach.
 We show that the expected leading-logarithmic terms emerge from our calculations along with recovering the proper argument of the running coupling in the soft limit and the  constant $K$ related to the emergence of the CMW or ``physical'' coupling \cite{Catani:1990rr,Banfi:2018mcq,Catani:2019rvy}. Apart from these standard ingredients we also derive the collinear NLL terms finding agreement with the SCET results \cite{Frye:2016aiz,Bell:2018vaa,Kardos:2020ppl}. Given our derivation directly from the QCD matrix-elements and independent of any previous SCET work, this constitutes a strong completely independent check on the main results involved in the NLL resummation for mMDT jet mass.  We shall also shed further light on the collinear structure we find here in terms of standard QCD resummation elements
 which should, we hope, enable the resummation of the mMDT jet mass also within a QCD resummation formalism e.g. through a suitable modification of the NNLL ARES method \cite{Banfi:2014sua,Banfi:2018mcq}  or directly by embedding the triple-collinear splitting within a strongly-ordered parton cascade. We note that calculations based on triple-collinear splitting functions in the context of jet physics have also been carried out in \\ Refs.~\cite{Alioli:2013hba,Bertolini:2015pka}. For early work on integrals of triple-collinear splittings in the context of initial state space-like splittings we refer the reader to Ref.~\cite{deFlorian:2001zd}.

 Although we carry out our present calculation in the context of $e^{+}e^{-}$ annihilation precisely as in Ref.~\cite{Frye:2016aiz} to control process dependent pieces of the calculation, the final NLL results we extract derive purely from the collinear limit and hence are universal. We confine our attention here to quark initiated jets though similar calculations can be performed for gluon jets. Also, while for the purposes of the present paper, we shall work in the small $\zc$ approximation purely for simplicity, we stress that our approach can be extended to obtain the finite $\zc$ corrections at the NLL level. Although finite $\zc$ NLL corrections are expected to be numerically small they should be of increased relevance in the context of the recent NNLL calculation performed in the small $\zc$ limit \cite{Kardos:2020gty}.

The current paper is organised as follows. We start in section \ref{sec:def+LL} by defining the normalised jet mass observable $\rho$, for which we use the same $e^{+}e^{-}$ definition as in Ref.~\cite{Frye:2016aiz} i.e. study the groomed heavy hemisphere mass, and write down the leading-logarithmic resummed formula with running CMW coupling. In section \ref{sec:lo} we compute the leading-order result in the small $\rho$ limit and identify the $\rho$ independent constant term $C_1 \, \alpha_s$ that multiplies the Sudakov factor in the LL formula. We then expand the LL result to order $\alpha_s^2$ and note that while this expansion also produces terms that are NLL in $\rho$, such terms originate within an LL context, being still related to the strongly-ordered picture. In section \ref{sec:secondorder} we carry out our $\alpha_s^2$ calculations for the jet mass differential distribution, $\rho \frac{d\Sigma}{d\rho}$, considering various real-emission configurations including emissions in opposite hemispheres and those in the same hemisphere. We compute separately the contributions from gluon emission with a $C_F^2$ colour factor and from gluon decay with $C_F C_A$, $C_F T_R n_f$ and $C_F (C_F-C_A/2)$ terms and combine the divergent real-emission results with those from virtual corrections. At this stage we compare our results to the expansion of the LL formula and identify the process-independent NLL terms. In section \ref{sec:discussion} we compare our results for each colour channel to the SCET results finding agreement in each case. We finally comment on the nature of our results within the QCD resummation context and briefly mention some prospects for further work. Various formulae relevant to the derivation of our results are listed in the appendices.

\section{Observable definition and leading-log resummation
\label{sec:def+LL}}

We are interested in the jet mass distribution of a QCD  jet after the application of grooming via the modified mass-drop tagger (mMDT) or equivalently SoftDrop ($\beta=0$).  In the current article, for reasons of convenience, we shall work in the context of $e^{+} e^{-}$ collisions, though the NLL pure collinear terms we will eventually extract are process independent and hence apply also to jets at hadron colliders. %

We will compute the standard heavy hemisphere jet mass observable extensively studied in $e^{+}e^{-}$ collisions but with the modification that we compute the heavy jet mass after running the mMDT procedure on the particles in each hemisphere. This was also the observable studied in the first mMDT NLL resummed calculation performed in Ref.~\cite{Frye:2016aiz}. %
In general one may separate the event into two hemispheres in different ways e.g. by clustering to two jets as in Ref.~\cite{Frye:2016aiz} or as is traditional by using the thrust axis. At the level of our calculations and for extracting the terms we seek, we are insensitive to the precise details since in the soft and/or collinear limit we can take the hemispheres to be defined by the directions of the initial quark--anti-quark pair.

The mMDT algorithm, as applied to $e^{+}e^{-}$ collisions, involves declustering particles in a hemisphere of an $e^{+} e^{-}$  event using the Cambridge/Aachen (C/A) algorithm, as one does for a hadron collider jet. We produce, at each stage, two branches $i$ and $j$ and we require
\begin{equation}
\frac{\mathrm{\min}(E_i, E_j)}{E_i+E_j} > z_{\mathrm{cut}}.
\end{equation}
If a branching fails this requirement we reject the softer branch and proceed with declustering the harder branch until the condition passes or we end up with a massless hard parton.  We then define
\begin{equation}
\rho = \frac{\mathrm{max}(M_R^2,M_L^2)}{Q^2/4},
\end{equation}
where we select the larger of the left or right hemisphere squared invariant masses ($M_L^2$ and $M_R^2$ respectively) and normalise to $\left(Q/2  \right)^2$ which corresponds to the squared energy  of a hemisphere in the Born limit.

We shall work in the formal limit $\rho \ll \zc \ll1$, which means that we will examine the structure of $\ln \rho$ enhanced terms, but shall neglect power corrections in $\zc$ therin. A leading log (LL) resummation formula for the mMDT jet mass distribution, based on an independent emission picture with emissions strongly ordered in angles, was first provided in Ref.~\cite{Dasgupta:2013ihk} (see Eq.~(7.2) of Ref.~\cite{Dasgupta:2013ihk} for the result). While that result applies directly to a jet  produced in hadron collisions, it can be easily modified to the case of the heavy groomed hemisphere mass in $e^{+}e^{-}$ annihilation.  We express our leading-logarithmic resummed result in terms  of the integrated distribution, $\Sigma(\rho) =\int_0^{\rho} \frac{1}{\sigma_0} \frac{d\sigma}{\sd \rho'}  \sd \rho'$ as
\begin{eqnarray}
&&\Sigma^{\mathrm{LL}}(\rho) = \left(1+\frac {C_F \alpha_s}{2\pi}  C_1 \right)  \exp \left[-\int_{\rho}^{1} \frac{\sd \rho'}{\rho'}
\right.
\nonumber\\
&&\left.\int_0^{1-\mathrm{max} \left(z_{\mathrm{cut}}, \rho' \right)} \sd  z \, p_{qq} (z)
\frac{ C_F}{\pi}  \alpha_s^\CMW \left(\left(1-z \right) \rho' Q^2/4 \right) \right],
\label{eq:llresult}
\end{eqnarray}
where we defined the splitting function \\ $p_{qq}(z) = (1 + z^2)/(1-z)$ and $\sigma_0$ is the Born cross section.

The above result modifies the result of Ref.~\cite{Dasgupta:2013ihk} by replacing $p_T^2$ by $Q^2/4$ and is written in terms of the splitting function $p_{qq}(z)$ rather than
$p_{gq}(z)$ so that the variable $1-z$ in the above result corresponds to the variable $z$ in Eq.~(7.2) of Ref.~\cite{Dasgupta:2013ihk}. We have also inserted an additional  factor of two in the Sudakov exponent which takes into account the fact that we are considering the result due to two hard partons in the left  and right hemispheres respectively, rather than just  a single parton initiating a jet. This is accounted for by our definition of the splitting function which has a factor of two relative to that defined in Ref.~\cite{Dasgupta:2013ihk}. The above result is labelled LL since it correctly resums terms $\alpha_s^n L^n$, with $L= \ln \rho$, i.e. the leading logarithms in $\rho$. Additionally a fixed-coupling calculation of the exponent in Eq.~\eqref{eq:llresult} reveals a term of the form $\alpha_s \ln^2 \zc$, which although subleading in $\rho$, correctly resums terms $\left(\alpha_s \ln^2  \zc\right)^n$\cite{Dasgupta:2013ihk}.

In Eq.~\eqref{eq:llresult} we have defined the coupling in the CMW or ``physical" scheme \cite{Catani:1990rr,Banfi:2018mcq, Catani:2019rvy} and have set the argument of the running coupling to be the transverse momentum squared of a soft and collinear emission relative to the direction of its hard parent parton, expressed in terms of the invariant mass $\rho'$ and the energy fraction of the emitted gluon $\frac{Q}{2} (1-z)$,  since we have $k_t^2 = (1-z) \rho' Q^2/4.$ \footnote{Recall that in the soft-collinear limit $\rho' =m^2/(Q^2/4) \approx (1-z) \theta^2$ while $k_t = \frac{Q}{2} (1-z) \theta$.}
Both the specification of the CMW scheme and the precise details of the argument of the running coupling (beyond the fact that it scales with $\rho$) in fact produce terms that are formally NLL in $\rho$ but with logarithmic enhancements in $z_{\mathrm{cut}}$. These terms are an intrinsic part of the leading-logarithmic resummation framework and hence naturally belong in our LL formula.  In a similar spirit we have also included in Eq.~\eqref{eq:llresult}, a $\rho$ independent constant order $\alpha_s$ coefficient $C_1$, which is process dependent and on physical grounds factorises from the leading-logarithmic Sudakov resummation. On expanding the exponent, multiplication by the $C_1$ term results in an $\alpha_s^2 L$ term. While this term should be reproduced by our order $\alpha_s^2$ calculations, it is process dependent and unrelated to the collinear NLL structure we ultimately seek to extract. Hence the explicit identification of $C_1$ is needed, to account for the role of this piece in our final result.

We then expect that our eventual order $\alpha_s^2$ result should contain all terms generated by the expansion of \eqref{eq:llresult} and additionally all terms of genuine NLL origin (i.e. unrelated to the strongly-ordered in angle LL dynamics). These terms should arise from collinear physics, and be independent of $z_{\mathrm{cut}}$ and of the specific hard process.

\section{Leading order calculation and expansion of LL result }

Throughout this article we shall work in the resummation region where $\rho \ll z_{\mathrm{cut}}$ and hence ignore the presence of a transition point at $\rho \approx \zc$, beyond which the groomed jet mass result becomes coincident with the plain ungroomed mass. We shall consider for simplicity that $z_{\mathrm{cut}} \ll 1$, so that we can neglect powers of  $z_{\mathrm{cut}}$, although this is not a requirement on the validity of our triple-collinear calculations.

Our first step will be to determine the $\rho$ independent coefficient $C_1$ that appears in Eq.~\eqref{eq:llresult}. We do this in the following subsection by performing a leading-order calculation in the soft and collinear limit.

\subsection{Leading-order calculation \label{sec:lo}}

Here we derive, in the small $\rho$ and $\zc$ limit as defined before, the order $\alpha_s$ contribution to $\Sigma(\rho)$. At this order we have to consider a single real emission and the one-loop virtual correction  to $q\bar{q}$  production.  To handle divergences in the real emission calculation, we perform the calculation in conventional dimensional regularisation (CDR) in $d =4-2\epsilon$ dimensions, and combine with the virtual correction before taking $\epsilon \to 0$.

Since we are interested in the region of small jet mass and small $z_{\mathrm{cut}}$, we can work in the soft and/or collinear limit, perform the calculation for a single hemisphere and then double the result. Considering the gluon to be in the same hemisphere as the quark (for example) and applying the \\ mMDT, we have a situation where the $z_{\mathrm{cut}}$ condition passes or fails. If it passes then one obtains a non-zero hemisphere mass, while if it fails the hemisphere mass vanishes.

When the $z_{\mathrm{cut}}$ passes, we are considering a pair of relatively energetic partons which produce a small jet mass, implying that the angle between partons is small,  $\theta^2 \sim \frac{\rho}{z_{\mathrm{cut}}} \ll 1$, which allows the use of the collinear approximation. In this region the emission probability is just the $q \to q g$ splitting function in $4-2\epsilon$ dimensions $p_{qq}(z,\epsilon) = p_{qq}(z) -\epsilon(1-z) $, and the result follows by applying the standard d-dimensional collinear phase space \cite{Giele:1991vf}:

\begin{eqnarray}\label{eq:locoll}
 &&\Sigma_{1,\mathrm{r_1}}(\rho)  = 2\times\frac{ C_F \alpha_s}{2\pi}  \frac{(4\pi \mu^2)^\epsilon}{\Gamma(1-\epsilon)} \left(\frac{Q}{2}\right)^{-2 \epsilon}\int\frac{\sd\theta^2}{\theta^{2(1+\epsilon)}}
 \nonumber\\
 &&    \int_{\zc}^{1-\zc}\left(z(1-z) \right)^{-2\epsilon}
 p_{qq}(z,\epsilon)\,  \,\,  \Theta(z(1-z)\theta^{2}<\rho) \sd z,
\end{eqnarray}
where the label $r_1$  indicates the first of three distinct real-emission terms that we encounter, and we have provided an explicit factor of 2 to take account of both hemispheres, which shall apply to  all the real-emission terms we compute in this section. The constraint arising from restricting the jet mass is written in terms of a step function involving $\rho$, while the $\zc$ condition removes any soft singularity in the integral and we obtain just a collinear pole alongside finite corrections. We have not explicitly written, for brevity, the dependence on  $\zc$ in the argument of $\Sigma_{1,\mathrm{r_1}}$ which is left implicit here and throughout the article.

Next we turn to the region where the $z_{\mathrm{cut}}$ condition fails, the softer particle is groomed away, and we obtain a massless hemisphere. This can happen when the quark goes soft for $z<z_{\mathrm{cut}}$ or when the gluon goes soft for $z>1-z_{\mathrm{cut}}$, and in either case the jet mass vanishes so that we can replace the step function in Eq.~\eqref{eq:locoll} by unity. The region corresponding to a soft quark also contributes only a collinear divergence and finite terms which are suppressed by powers of $z_{\mathrm{cut}}$  which we neglect.  Hence retaining just the collinear divergence, corresponding to a pole in $\epsilon$ produced by the $\theta^2$ integral, we can write this contribution as
\begin{equation}
\label{eq:coll-2}
 \Sigma_{1,\mathrm{r_2}} (\rho)  = -2 \frac{ C_F \alpha_s}{2\pi} \frac{1}{\epsilon}
 \int_0^{z_{\text{cut}}} \left(\frac{1+z^2}{1-z} \right)  \, \sd z \ \ ,
\end{equation}
where we performed the  $\theta^2$ integral and discarded all finite terms owing to their suppression with $z_{\mathrm{cut}}$. Accordingly we have dropped the $\epsilon$ dependence of the constant prefactor multiplying the integral, since this only leads to finite power suppressed in $\zc$ corrections.

Lastly there is the region $z>1-z_{\mathrm{cut}}$ corresponding to a soft gluon emission, which is removed by grooming. As there is no longer a constraint on the angle, one has to replace the soft-enhanced part of the splitting function with the full eikonal function to account for the wide-angle region. For soft-regular pieces, however, one merely needs the collinear pole as any finite contributions are power suppressed in $\zc$. Hence, using $p_{qq}(z,\epsilon) =
2/(1-z)-(1+z)-\epsilon (1-z)$, to separate the soft enhanced and regular terms of the splitting function, we write
\begin{align}
 \Sigma_{1,r_3} =  \Sigma_{1,r_3 \, \mathrm{soft}} + \Sigma_{1,r_3 \, \mathrm{coll.}} \ \ ,
 \end{align}
where explicitly
\begin{multline}
\label{eq:soft-coll}
\Sigma_{1,r_3 \, \mathrm{soft}} =  2\frac{ C_F \alpha_s}{2\pi}  \frac{(4\pi \mu^2)^\epsilon}{\Gamma(1-\epsilon)} \left(\frac{Q}{2}\right)^{-2 \epsilon}
\times \\
\int_{1-z_{\mathrm{cut}}}^{1}   \frac{2\, \sd z }{(1-z)^{1+2\epsilon}}  \,
\int_0^1 \frac{2   \  \sd \left(\cos \theta \right)}{\left(1-\cos\theta \right)^{1+\epsilon} \left(1+\cos\theta \right)^{1+\epsilon}}\ \ ,
\end{multline}
\begin{equation}
\Sigma_{1,r_3 \, \mathrm{coll.}} = 2\times\frac{ C_F \alpha_s}{2\pi} \frac{1}{\epsilon}\int_{1-z_{\mathrm{cut}}}^{1} \sd z (1+z)\ .
\end{equation}
In the expression for $\Sigma_{1,r_3 \, \mathrm{soft}}$ the  upper limit of the $\cos \theta$ integral corresponds to $\theta=\pi/2$ for the boundary of the hemisphere under consideration, while in the expression for  $\Sigma_{1,r_3 \, \mathrm{coll.}}$ we have retained only the singular contribution as finite corrections vanish with $\zc$.

Our final ingredient is the well-known virtual correction to $q\bar{q}$ production (see e.g. \cite{Giele:1991vf}):
\begin{eqnarray}
\label{eq:virtborn}
&&\mathcal{V}(\epsilon) = \frac{C_F \alpha_s}{2\pi} \frac{\Gamma(1+\epsilon) \Gamma^2 (1-\epsilon)}{\Gamma(1-2\epsilon)}\times
\nonumber\\
&&\hspace{1.0cm}\left( -\frac{2}{\epsilon^2} \left(\frac{4 \pi \mu^2}{Q^2} \right)^\epsilon +\pi^2-8 -\frac{3}{\epsilon}
\left( \frac{4 \pi \mu^2}{Q^2} \right)^\epsilon \right).
\end{eqnarray}
We express our results in terms of the renormalised coupling in the $\overline{\mathrm{MS}}$ scheme, $\alpha_s(\mu_R^2)$, using the relation
\begin{equation}
\label{eq:renorm}
\mu^{2\epsilon} \alpha_s = S^{-1}_\epsilon\mu_R^{2\epsilon} \alpha_s(\mu_R^2) +\mathcal{O}(\alpha_s^2),
\end{equation}
where  we have the standard $\overline{\mathrm{MS}}$ factor
\begin{equation}
\label{eq:msbar}
S_\epsilon = (4\pi)^\epsilon e^{-\epsilon \gamma_E},
\end{equation}
and we choose $\mu_R =Q/2$. \footnote{Beyond order $\alpha_s$,  the RHS of Eq.~\eqref{eq:renorm} contains UV poles in $\epsilon$ related to the renormalisation of the strong coupling.}

Carrying out the necessary integrals we obtain the result for the real emission term $\Sigma_{1,r} = \Sigma_{1,r_1}+\Sigma_{1,r_2}+\Sigma_{1,r_3  \,  \mathrm{soft}}+ \Sigma_{1,r_3 \, \mathrm{coll.}}$,
\begin{equation}
\label{eq:realres}
\begin{split}
\Sigma_{1,r}  = \frac{C_F \alpha_s \left(Q^2/4 \right)}{2\pi} & \bigg( \frac{2}{\epsilon^2} + \frac{3-4 \ln 2}{\epsilon} -\ln  \rho\left(4\ln \zc+3 \right) \\
&+2 \ln^2 \zc + 8 \ln 2 \ln \zc\\
&+4\ln^2 2 -\frac{7 \pi^2}{6}+7 \bigg).
\end{split}
\end{equation}
Combining with the virtual correction we obtain the leading-order result, $\Sigma_1(\rho) =\Sigma_{1,r}+\mathcal{V}(\epsilon)$:
\begin{equation}\label{eq:LO}
\begin{split}
\Sigma_1 (\rho)= \frac{C_F \alpha_s \left(Q^2/4\right)}{2\pi} & \bigg[\left( -\ln \frac{1}{\rho} \left(4 \ln \frac{1}{\zc}-3\right)+2\ln^2\zc \right) \\
&- 2 \ln 2\left( 4 \ln \frac{1}{\zc}-3\right) -1\bigg],
 \end{split}
\end{equation}
where the  argument of the running coupling reflects our choice of $\mu_R$. We have written the above result separating the contributions that arise from expanding the LL formula (i.e the $\ln 1/\rho$ and $\ln^2 \zc$ contributions that are associated to the fixed-coupling Sudakov exponent) from the contributions we shall associate to $C_1$. This allows us to identify
\begin{equation}
\label{eq:constant}
C_1 =  -2 \ln 2 \left( 4 \ln \frac{1}{\zc}-3\right) -1.
\end{equation}
We note that defining the observable as $v=\rho/4$, corresponding to a normalisation to $Q^2$, would result in the elimination of the term proportional to $\ln2$ and hence the $\ln z_{\mathrm{cut}}$ dependence from $C_1$, but the resummation of $\ln \zc$ terms is beyond the scope of our results.

\subsection{LL result at order $\alpha_s^2$}

On performing the full order $\alpha_s^2$ calculation, to our NLL in $\rho$ accuracy, we should recover all terms produced by the expansion of Eq.~\eqref{eq:llresult} in addition to terms that are unrelated to the leading logarithmic structure, which will then act as one of the checks on our results. To this end we report below the expansion of  Eq.~\eqref{eq:llresult} for the differential distribution $\rho \frac{\sd\Sigma}{\sd\rho}$ (to correspond to the calculations of the following sections). The leading-order result is given by (for our choice of $\mu_R =Q/2$),
\begin{equation}
\rho \frac{\sd\Sigma^{\mathrm{LL}}_1}{\sd \rho} = \frac{C_F  \alpha_s\left( Q^2/4 \right)}{2\pi}  \left( -3-4 \ln \zc \right).
\end{equation}
To obtain the order $\alpha_s^2$ result one needs to compute the Sudakov exponent with running coupling and switch from the CMW scheme to the $\overline{\mathrm{MS}}$ scheme, which gives
\begin{equation}\label{eq:llexpansion}
\begin{split}
\rho \frac{\sd\Sigma^{\mathrm{LL}}_2}{\sd \rho} =  \left(\frac{\alpha_s}{2\pi}\right)^2 &\bigg[  C_F^2\bigg((3+4\ln\zc)^2\ln\rho -8 \ln ^3\zc
\\
 &- 2 (3+16 \ln 2) \ln ^2\zc \\
 &+ (4-48 \ln2) \ln \zc -18 \ln2+3 \bigg) \\
 &+  C_F C_A \left(  \frac{11}{6} \left( 3+4\ln\zc\right)\ln\rho \right.\\
 & \left.+\frac{11}{3} \ln^2 \zc +\left(\frac{2\pi^2}{3}-\frac{134}{9} \right) \ln \zc \right)  \\
 &+ C_F T_R n_f \left( -\frac{2}{3}(3+4\ln\zc)\ln\rho\right.
 \\
 & \left.-\frac{4}{3}\ln^2\zc  + \frac{40}{9} \ln\zc \right) \bigg].
\end{split}
\end{equation}
The above result contains a term which is LL in $\rho$ originating from the exponentiation of the leading-order result. It also contains NLL in $\rho$ terms, corresponding to $\rho$ independent terms in $\rho \, \sd \Sigma_2/ \sd \rho$ generated by an interplay of the LL exponent with $C_1$ and by fixing the scale and scheme of the running coupling. In particular the $\ln^2  \zc$ terms in the $C_F C_A$ and $C_F T_R n_f$ channels derive from the $1-z$ factor in the argument of the running coupling while the $\ln \zc$ term in the same channels is generated by changing from the CMW scheme to the $\overline{\mathrm{MS}}$ scheme as can be seen through their coefficient, proportional to $K = \left( \frac{67}{18}-\frac{\pi^2}{6} \right) C_A -\frac{10}{9} T_R n_f $.
In addition to these terms, the $1-z$ factor in the argument of the running coupling and the CMW coefficient $
K$ applied to the full splitting function, rather than just its soft enhanced piece, are also responsible for producing $z_{\mathrm{cut}}$ independent NLL terms, which go beyond the strict jurisdiction of the LL formula. The full set of such terms will be identified through the calculation we perform here and can be properly accommodated within a consistent NLL resummation formula.

In the next sections we shall derive the full result at order $\alpha_s^2$ through to NLL accuracy, compare it to the expectations from Eq.~\eqref{eq:llexpansion} and derive the $\zc$ independent NLL corrections.

\section{NLL at $\mathcal{O} \left (\alpha_s^2 \right)$: the triple-collinear limit\label{sec:secondorder}}

At LL accuracy, for the mMDT jet  mass, we have a picture of successive collinear parton branchings which are strongly ordered in angle with each branching being described by a leading-order (LO) $1 \to 2$ splitting function. Thus at order $\alpha_s^2$, the real emission matrix-element simply involves a product of two LO splitting functions. To obtain NLL accuracy, at order  $\alpha_s^2$, one has to consider three partons that are comparably collinear i.e. the opening angle between any two partons is small $\theta_{ij}^2 \ll 1$ but there is no strong ordering so that $\theta_{12}^2 \sim \theta_{13}^2 \sim \theta_{23}^2$. Such configurations are described by the $1 \to 3$ collinear splitting of an initial parton, and the matrix-element involves triple-collinear splitting functions. In the  strongly-ordered limit, the triple-collinear splitting functions reduce to a product of LO splitting functions (in general after azimuthal averaging) thus restoring the LL picture.

For our current calculations, the relevant functions are the unpolarised triple-collinear splitting functions for a quark (or anti-quark) initiated $1 \to 3$ splitting, denoted $\langle P_{abq}  \rangle$ for a splitting $q \to q a b$ with $a$ and $b$ representing parton flavours, which were first computed in Refs.~\cite{Campbell:1997hg,Catani:1998nv,Catani:1999ss}, and are listed in the appendix.
For the $q \to ggq$ splittings there is both a gluon emission contribution with a $C_F^2$ colour factor, and a gluon decay contribution with a $C_F C_A$ colour factor. The $q \to q \bar{q} q$ splitting arises from gluon decay and has  a $C_F  T_R n_f$ colour factor as well as a contribution from an identical particle interference contribution, involving quarks of the same flavour in the final state, which has a colour factor $C_F \left(C_F-C_A/2 \right)$ i.e. vanishes in the leading $N_c$ limit. Identical considerations apply for the decay of an initial anti-quark. In the following subsections, we consider the gluon emission and decay contributions in turn.

\subsection{Gluon emission contribution}

Here we study the emission of two gluons from the initial $q \bar{q}$ system, associated with a $C_F^2$ colour factor.
The emitted gluons can either be in the same or in opposite hemispheres, with the latter case being simply related to the leading-order calculations we have already performed. We deal with each contribution in turn below.

\subsubsection{Emissions in opposite hemispheres}

Consider a gluon emitted in each of the ``right" and ``left'' hemispheres containing the quark and anti-quark respectively. Let us assume that the right hemisphere is heavier after grooming and that its groomed jet mass is $\rho$. This implies that the branching in the right hemisphere must pass the $\zc$ requirement corresponding to $1-\zc > z > \zc$ and that it must set a mass $\rho$, while in the left hemisphere the mass must be below $\rho$ for it to be lighter, and hence the grooming can either retain both or remove one of the two particles. The fact that in the right hemisphere the grooming passes, coupled with the limit we are working in, with $\rho \ll \zc$, allows us to use the collinear approximation, so that the branching in the right hemisphere factorises from the dynamics of the left hemisphere. Here we do not require the triple-collinear splitting functions, as the emissions in opposite hemispheres are well separated.

For the left hemisphere the constraint on the mass to be below $\rho$ simply gives us $\Sigma_{1,r}/2$ with $\Sigma_{1,r}$ the real emission result already computed in the previous section (see Eq.~\eqref{eq:realres}. For the right hemisphere the distribution can be simply calculated in the collinear limit using the LO splitting function and the collinear $1 \to 2$ phase space. Finally a factor of two accounts for the case when the left hemisphere is heavier after grooming.

For compactness, here and in the sections below, we define the quantity  $\mathcal{F}(\rho)$,
\begin{equation}
\mathcal{F}(\rho) = \rho \frac{d\Sigma_2}{d\rho},
\end{equation}
where $\Sigma_2$ is the order $\alpha_s^2$ contribution to $\Sigma(\rho)$.
The result for the emissions in opposite hemispheres can then be written as a product of two leading-order factors\footnote{Having specified our choice of $\mu_R =Q/2$ we shall not explicitly indicate the argument of $\alpha_s$ in the order $\alpha_s^2$ pieces.}:
\begin{eqnarray}
&&\mathcal{F}^{\mathrm{opp.}}(\rho,\epsilon) = \Sigma_{1,r} \times \frac{C_F \alpha_s}{2\pi} \frac{e^{\epsilon \gamma_E}}{\Gamma(1-\epsilon)}
\times\nonumber\\
&&\hspace{2cm}\int_{\zc}^{1-\zc} \left (\rho z (1-z) \right)^{-\epsilon} p_{qq} \left(z,\epsilon\right)  \sd z.
\end{eqnarray}
Note that although the LO jet mass distribution for the right hemisphere is a finite quantity, we have retained its $\epsilon$ dependence in the result above, since
$\Sigma_{1,r}$ contains double and single $\epsilon$ poles and finite terms (in the limit $\epsilon  \to 0$) are generated by the $\epsilon$ expansion.

\subsubsection{Emissions in the same hemisphere}

\begin{figure}[h]
\centering
 \includegraphics[trim = 0 0 0 0,clip=true, width= 0.45\textwidth]{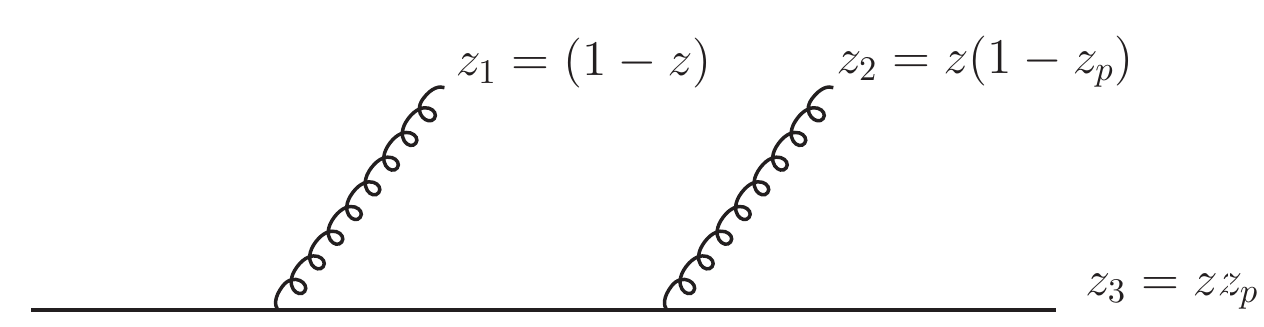}
 \caption{An illustration of the parameterisation used for kinematic variables
 in the gluon emission process, relevant to the triple-collinear limit calculation discussed in the main text.}
\label{fig:CFCF}
\end{figure}

When two emissions, i.e. three partons, are in the same hemisphere one has to consider the action of the mMDT taking into proper account the Cambridge/Aachen clustering sequence. This involves considering different angular regions where the two emissions can be clustered separately to the quark (or anti-quark according to the hemisphere in question) or are first clustered together and then clustered to the quark. It proves convenient to divide the calculation into two pieces: a first piece where in all angular regions we apply the mMDT as if the gluons are clustered separately to the quark and a second piece which restores the correct action of the mMDT in the angular region where the two gluons are clustered. We are then led to consider the following distinct cases:

\begin{itemize}

\item \underline{Larger-angle gluon passes $\zc$}: Neglecting the clustering of the two gluons, the mMDT declustering produces two branches, consisting of the larger-angle gluon and a massive branch with the quark and the smaller-angle gluon. When the first declustering passes the $\zc$ condition all three partons are retained. The angle between the branches is small, being set by $\theta^2 \sim  \rho/\zc$ and hence the three partons are constrained to be within a small angular region and the triple-collinear limit generates the full result. The result for the real emission calculation will be divergent due to the smaller-angle emission becoming soft and/or collinear and will contain double and single $\epsilon$ poles.

\item \underline{Larger-angle gluon fails $\zc$} : Another relevant situation is that the larger-angle gluon is soft and hence the first declustering fails the $\zc$ condition.\footnote{The first declustering can also fail the $\zc$ condition due to the massive quark-gluon branch going soft, but rejection of this branch leads to a massless hemisphere and hence this contribution can be ignored for $\rho \neq$ 0.} In this case the soft gluon is groomed away while the tagger then declusters the second gluon and we require the second declustering to pass the  $\zc$ condition to obtain a massive hemisphere. In this case only the smaller-angle gluon is constrained by the jet mass to be collinear to the quark, while the first emission can be at a large angle. Hence we need to modify the triple-collinear splitting functions to match the correct soft large-angle emission pattern for the first gluon, precisely as we did in section \ref{sec:lo}. The result is divergent due to the soft divergence produced by the larger-angle gluon, though it contains only a single pole in $\epsilon$.

\item \underline{Correction for gluon clustering}: In the region where the angle between the two gluons, $\theta_{12}$, is the smallest angle the gluons are clustered in the C/A algorithm. Hence the first declustering produces a massive branch with the two gluons and a massless branch i.e. the quark. If the quark is soft the $\zc$ condition can fail and the tagger recurses down the massive two-gluon branch. However such configurations with a soft quark are suppressed by powers of $\zc$ and can be ignored consistent with our intended accuracy. The two-gluon branch must also pass the $\zc$ as grooming it away would lead to a massless jet. Hence we can always take the first declustering to pass the $\zc$ condition so that all three partons are retained, implying that  the triple-collinear limit is once again the relevant one. In the angular region where the gluons are clustered, we shall {\emph{subtract}} the contributions already included in the first two scenarios described above, and shall add the correct constraints just discussed. The difference between the correct and subtraction terms is finite and can be calculated in four dimensions.
\end{itemize}

For calculations in the triple-collinear limit,  we work in terms of the energy fractions $z_i$, defined wrt the initiating parton's energy, and which satisfy $\sum_{i} z_i=1$, and the angles $\theta_{ij}$ between any two partons $i$ and $j$, such that $\theta_{ij}  \ll1$. The triple-collinear phase space in $4-2\epsilon$ dimensions may be written as \footnote{Additionally a $1/2!$ symmetry factor applies for identical particles in final state.}
\cite{Gehrmann_De_Ridder_1998}
\begin{eqnarray}
\label{eq:tcphasesp}
&&\text{d} \Phi_3 = \frac{1}{\pi}\left(\frac{Q}{2}\right)^{4-4\epsilon} \frac{1}{\left(4\pi\right)^{4-2\epsilon} \Gamma\left(1-2\epsilon\right)}
 \times\nonumber\\
&&\hspace{.5cm}\text{d}z_2 \text{d}z_3  \text{d} \theta_{13}^2    \text{d} \theta_{23}^2
\text{d} \theta_{12}^2\, (z_1 z_2 z_3)^{1-2\epsilon}\, \Delta^{-1/2-\epsilon}  \, \Theta (\Delta),
\end{eqnarray}
where the Gram determinant $\Delta$ is defined as
\begin{equation}
\Delta = 4 \theta_{13}^2 \theta_{23}^2 - \left(\theta_{12}^2-\theta_{23}^2-\theta_{13}^2\right)^2.
\end{equation}

To make contact with the LO splitting functions in the strongly-ordered limit, it is useful to parameterise the variables $z_i$ in terms of variables $z$ and $z_p$ as depicted in Fig.~\ref{fig:CFCF}. Our general method for integrating the triple-collinear functions and extracting its divergences is briefly described in \ref{sec:integrals}. It involves systematic subtraction of soft and collinear divergences, via a series expansion around divergent limits, to obtain the pole structure and a pure finite contribution which we integrate numerically in four dimensions. Our results shall thus be partly analytical (stemming from performing an $\epsilon$ expansion of the coefficients of the pole terms) and partly numerical in nature.

We first provide the details for results neglecting the clustering of gluons starting from the contribution where the larger-angle emission passes the $\zc$, which we label $\mathcal{F}^{\mathrm{pass}} (\rho,\epsilon)$. We take
$\theta_{13}$ to be the larger angle and hence for the first declustering to pass the $\zc$ condition we have that $1-\zc> z> \zc$. The smaller angle gluon is not examined for the $\zc$ condition and can be arbitrarily soft and/or collinear leading to divergences in both limits i.e. from $z_p \to 1$ and $\theta_{23} \to 0$. The relevant splitting function and prefactor is specified by Eqs. \eqref{eq:tripcollfac}, \eqref{eq:qggabsf} and we obtain
\begin{eqnarray}
\label{eq:zpass}
&&\mathcal{F}^{\text{pass}}(\rho,\epsilon)= 2 S_\epsilon^{-2} \left( \frac{Q}{2} \right)^{4\epsilon}
\times\nonumber\\
&&\int \sd \Phi_3 \frac{\left(8 \pi \alpha_s \right)^2}{s_{123}^2}
C_F^2 \la P_{q\rightarrow g_1 g_2 q_3 }^{(ab)}\ra
\nonumber\\
&&\hspace{1cm}\delta_\rho(1,2,3)\Theta_{\zc}(1|23) \Theta(\theta_{23}<\theta_{13}),
\end{eqnarray}
where $s_{123}^2 = \frac{Q^2}{4} \sum_{j >i}^{3}  \sum_{i=1}^{2}z_i z_j \theta_{ij}^2$ is the squared invariant mass of the three parton system and $\delta_\rho(1,2,3)$ is an abbreviated notation for the condition that
the normalised hemisphere jet mass $\rho$ involves all three partons i.e. the condition
\begin{equation}
\delta_\rho(1,2,3) =\rho \,\delta \left(\rho - \frac{4 s_{123}^2}{Q^2} \right),
\end{equation}
where the factor of $\rho$ in front of the delta function takes care of the fact that we are studying the logarithmic derivative $\rho  d\Sigma/d\rho$. We shall also use the notation $\Theta_{\zc}(a|b)$ to denote the condition that two branches $a$ and $b$, made up of one or two partons, pass the $\zc$ condition. Partons not included in $a$ and $b$ fail the $\zc$ condition and are removed by grooming. Thus $\Theta_{\zc}(1|23)$ in Eq.~\eqref{eq:zpass} indicates that both branches i.e. gluon $1$ and the massive branch with gluon $2$ and the quark (with index $3$) pass the $\zc$ condition. More explicitly we have  $\Theta_{\zc}(1|23) = \Theta \left(z<1-\zc \right)\Theta \left(z>\zc\right)$, amounting to simply a cut on $z$.

In Eq.~\eqref{eq:zpass} we have also introduced the renormalised $\overline{\mathrm{MS}}$ coupling $\alpha_s(\mu_R^2)$ via the use of Eq.~\eqref{eq:msbar} leading to the appearance of the $S_\epsilon^{-2}$ factor and chosen $\mu_R = Q/2$, though for brevity we have not explicitly written the argument of the running coupling above. We have introduced a factor of two to account for the other hemisphere containing the branching of the anti-quark.

Carrying out the integrals with the method discussed in the appendix, the result can be expressed in the following form:
\begin{equation}
\begin{split}
\mathcal{F}^{\text{pass}}(\rho,\epsilon) &= \left(\frac{C_F \alpha_s}{2\pi}\right)^2 \times\\
\int_{z_{\mathrm{cut}}}^{1-z_{\mathrm{cut}}} 2 &\bigg( \frac{H^{\text{soft-coll.}}(z,\rho,\epsilon)}{ \epsilon^2}
+ \frac{H^{\mathrm{coll.}}(z,\rho,\epsilon)}{\epsilon}\\
&+\frac{H^{\mathrm{soft}}(z,\rho,\epsilon)}{\epsilon} + H^{\text{fin.}}(z) \bigg) \sd z,
\end{split}
\end{equation}
where one notes the presence of a double pole coming from the soft $z_p \to 1$ and collinear $\theta^2_{23} \to 0$ limit and where single-pole contributions are separated into the contributions from soft ($z_p \to1$) and collinear ($\theta^2_{23} \to 0$) divergences alone given, respectively by the $H^{\mathrm{soft}}$ and $H^{\mathrm{coll.}}$ functions. We have
\begin{equation}
\begin{split}
H^{\text{soft-coll.}}(z,\rho,\epsilon) &= p_{qq}(z,\epsilon) z^{-2 \epsilon }\rho^{-2\epsilon}
\left(1-\frac{\pi^2}{6}\epsilon^2+\mathcal{O}(\epsilon^3)\right),\\
H^{\text{coll.}}(z,\rho,\epsilon) &= p_{qq}(z,\epsilon) z^{-2 \epsilon }\rho^{-2\epsilon}\times\\
&\left(\frac{3}{2}+\frac{13}{2}\epsilon-\frac{2\pi^2}{3}\epsilon+\mathcal{O}(\epsilon^2)\right),\\
H^{\text{soft}}(z,\rho,\epsilon) &= 0.
\end{split}
\end{equation}
The function $H^{\text{fin.}}(z)$ represents a finite contribution whose precise analytic form we have not extracted. Instead we study this finite contribution by  direct numerical integration over the triple-collinear phase-space in 4 dimensions. The result for the integration of $H^{\text{fin.}}(z)$ gives a constant as  $\zc \to 0$. The result that we obtain using integration with Suave \cite{Hahn_2005}, setting $\zc=0$ is,
\begin{equation}\label{eq:finitezpass}
2\int_{0}^{1} H^{\text{fin.}}(z) \sd z = 1.866 \pm 0.002.
\end{equation}
Next, we study the situation where the larger-angle gluon, i.e. emission 1, fails the $\zc$ condition and is groomed away, corresponding to $1>z>1-\zc$. This leaves the mass to be set by the smaller-angle emission 2, $\rho = z_2 z_3  \theta_{23}^2 = z^2 z_p(1-z_p) \theta_{23}^2$. This emission must survive grooming i.e.  $1-\zc  > z_p > \zc$ and hence $\theta_{23}^2 \ll 1$ for $\rho \ll \zc$.  The softness of emission 1, with energy proportional to $1-z$, implies that any terms regular in the limit $z \to 1$ produce power suppressed corrections in $\zc$ which we can neglect. Hence the only contribution comes from the singularity as $z \to 1$, which produces an $\epsilon$ pole and associated finite corrections.

We  start by considering the triple-collinear splitting function $P_{q\rightarrow g_1 g_2 q_3 }^{(ab)}$ and its integral over the phase-space, as for the previous case where emission 1 passes the $\zc$ condition. However since now emission $1$ fails the $\zc$ condition and is groomed away, it is not constrained to be collinear and has a range of angular integration going from $\theta_{13} \sim \theta_{23} \ll 1$ up to the boundary of the hemisphere at $\theta_{13} = \pi/2$. Near the lower limit of integration the triple-collinear approximation is valid, but to account correctly for soft emission at large angles we have to modify the angular dependence of the integral. This is precisely the same modification we made to account for soft large-angle emission for the calculation of $C_1$ (c.f. Eq.~\eqref{eq:soft-coll}). After neglecting pieces which contribute only an $\mathcal{O}(\zc)$ term on integration, we find the result
\begin{equation}\label{eq:ffail}
\begin{split}
\mathcal{F}^{\text{fail}}(\rho,\epsilon) &= \left(\frac{C_F \alpha_s}{2\pi} \right)^2\frac{2e^{2\epsilon\gamma_E}}{\Gamma(1-2\epsilon)}
\times\\
&\int_{\zc}^{1-\zc} \left((1-z_p) z_p\right)^{-2\epsilon} p_{qq}(z_p,\epsilon) \sd z_p
\\
&\int_{1-\zc}^1\left(1-z \right)^{-2\epsilon}\frac{2\ \sd z}{1-z}
\\
&\int_0^{1-\frac{\rho}{2 z_p(1-z_p)}} \frac{2 \ \sd \left(\cos \theta_{13} \right)}{\left(1-\cos\theta_{13} \right)^{1+\epsilon} \left(1+\cos\theta_{13} \right)^{1+\epsilon}}\
\\
&\int\frac{\sd \theta_{23}^2}{\theta_{23}^{2(1+\epsilon)}} \delta_\rho(2,3) \ ,
\end{split}
\end{equation}
where $\delta_\rho(2,3)$ is the condition that emission 2 and the quark labeled 3 contribute to the hemisphere invariant mass $\rho$. In fact, one can directly reach the same equation by realizing that the emission probability of a hard-collinear gluon completely factorizes from that of a soft gluon, i.e. the gluons are emitted independently in this region of phase space. Therefore, the total emission probability is a product of an eikonal function and a LO splitting function. This factorized structure is manifest in Eq.~\eqref{eq:ffail}.
The integral over $\theta_{23}^2$ is trivially performed using the delta function constraint which sets $\rho = z^2 z_p (1-z_p) \theta_{23}^2 \approx z_p (1-z_p) \theta_{23}^2$, where we have used the fact that  $z  \sim 1$ corresponding to the softness of emission 1.\footnote{Retaining the $z$ dependence in the mass constraint produces terms that vanish with $z_{\mathrm{cut}}$ and are beyond our accuracy.} We have modified the angular dependence so that at small $\theta_{13}$ we obtain the result arising from the triple-collinear splitting functions but for $\theta_{13} \sim 1$ we have the correct angular dependence for a  soft emission emitted off the $q\bar{q}$ dipole. We have also introduced the renormalised
$\overline{\mathrm{MS}}$ coupling and choose $\mu_R =Q/2$ as before. Evaluating the integrals we obtain:

\begin{equation}
\begin{split}
\mathcal{F}^{\text{fail}}(\rho,\epsilon)&=\left(\frac{C_F \alpha_s}{2\pi} \right)^2\times\\
&\int_{\zc}^{1-\zc} \sd z_p \, p_{qq}(z_p,\epsilon) \bigg[-\frac{2}{\epsilon} \ln \frac{4 z_p(1-z_p)}{\rho}
\\
&\hspace{2cm}-\frac{\pi^2}{3}+4\ln^2 2 -\ln^2 \frac{z_p(1-z_p)}{\rho}
\\
&\hspace{1cm} +2\ln \frac{4 z_p(1-z_p)}{\rho}\ln(\zc^2(1-z_p)z_p\rho) \bigg] \ \ .
\end{split}
\end{equation}
Finally we account for the correct action of the tagger when emissions $1$ and $2$ i.e. the two gluons are clustered first in the C/A algorithm and then the gluon pair is clustered to the quark. On applying the tagger one first encounters two branches, consisting of the quark and the massive gluon pair respectively.  If the quark fails the $z_{\mathrm{cut}}$ condition, one would then follow the branch consisting of the gluon pair. However such configurations with a soft quark are suppressed by powers of $z_{\mathrm{cut}}$ and hence ignored. On the other hand configurations where the massive gluon branch fails the $z_{\mathrm{cut}}$ condition would lead to a massless hemisphere.  Hence we only need to study the situation where both branches pass the $z_{\mathrm{cut}}$ condition and all three partons are retained. The opening angle between the branches is small, being set by $\rho/\zc$, which implies that all three partons are collinear and we can use purely triple-collinear kinematics.  To correct our earlier results, we simply need to calculate the difference between the correct configuration described here and our simplified treatment included as part of $\mathcal{F}^{\text{fail}}(\rho,\epsilon)$ and  $\mathcal{F}^{\text{pass}}(\rho,\epsilon)$. The relevant angular region for the calculation is $\theta_{12}^2 < \theta_{23}^2$, which corresponds to the C/A clustering of the two gluons, since we already have the ordering $\theta_{13}^2 > \theta_{23}^2$. Our clustering correction takes the form
\begin{equation}\label{eq:clustcf2}
\begin{split}
\mathcal{F}^{\text{clust.}}_{C_F^2} = \lim_{\zc\to0} 2  &\int \frac{\left(8\pi \alpha_s\right)^2}{s_{123}^2} \,C_F^2 \,
\la P_{q\rightarrow g_1 g_2 q_3 }^{(ab)}\ra \sd\Phi_3 \, \\
&\hspace{-1.5cm}\Theta(\theta_{23}<\theta_{13}) \Theta(\theta_{12}<\theta_{23})
\bigg( \delta_\rho(1,2,3) \Theta_{\zc}(3|12)  \\
&\hspace{-.5cm}- \delta_\rho(1,2,3) \Theta_{\zc}(1|23) - \delta_\rho(2,3) \Theta_{\zc}(2|3) \bigg),
\end{split}
\end{equation}
where in the second line the first term in parentheses represents the correct treatment of the tagger while the second and third terms correspond to the removal of the gluon clustering region from $\mathcal{F}^{\text{pass}}(\rho,\epsilon)$ and $\mathcal{F}^{\text{fail}}(\rho,\epsilon)$ respectively.

In the angular region relevant to their clustering, the smallest angle is that between the two gluons, so there is no collinear divergence in the gluon emission channel. We have potential soft divergences as each of $z$  and $z_p$ tend to 1  (i.e. $z_1$ or $z_2$ vanish), but in both those limits the correct calculation cancels with the subtraction terms in the second line of Eq.~\eqref{eq:clustcf2},  so that the result is purely finite and we can set $\epsilon =0$  in the integrals  that follow.

In the correct treatment, i.e. the first step function on the second line of Eq.~\eqref{eq:clustcf2}, the condition that the first declustering passes the $\zc$ corresponds to $1-\zc >z _3 >\zc$, where $z_3 = z z_p$. There are then two regions for the integration over energy fractions, according to the range of $z$:
\begin{equation}\label{eq:zlimits}
\begin{split}
&\zc<z<1-\zc, \quad \frac{\zc}{z}<z_p<1, \quad \text{and}
\\
&\quad 1-\zc<z<1, \quad \frac{\zc}{z}<z_p<\frac{1-\zc}{z} \ \ .
\end{split}
\end{equation}
For the region $\zc<z<1-\zc$, this corresponds to the region in $z$ where gluon 1 passes the $\zc$ in our simplified calculations, so we need to correct the calculation of $\mathcal{F}^{\text{pass}}(\rho,\epsilon)$ and the third term in the second line of  Eq.~\eqref{eq:clustcf2} vanishes. In the region $1>z>1-\zc$ corresponding to the grooming away of gluon 1 in the simplified calculation, we need to correct the treatment of $\mathcal{F}^{\text{fail}}(\rho,\epsilon)$ and here the second term in Eq.~\eqref{eq:clustcf2} vanishes.

In the former case one has a particularly simple situation since both in the simplified calculation and the correct treatment all three partons are retained and contribute to the jet mass via the $\delta_\rho(1,2,3)$ condition. This results in an identical angular integration in both cases and the difference between the correct and simplified treatment is purely due to the different limits on $z_p$. Denoting the result of the angular integrals by $\text{I}(z,z_p)$, we can write the clustering correction as:
\begin{equation}\label{eq:CfClustVanish}
\begin{split}
&\int_{\zc}^{1-\zc}\left(\int_{\frac{\zc}{z}}^{1}\text{I}(z,z_p) \sd z_p-\int_{0}^{1}\text{I}(z,z_p) \sd z_p\right) \sd z=
\\
&-\int_{\zc}^{1-\zc} \sd z\int_{0}^{\frac{\zc}{z}}\text{I}(z,z_p) \sd z_p,
\end{split}
\end{equation}
where the subtracted term above represents the removal of the simplified calculation in the gluon clustering region and where
\begin{equation}
\begin{split}
\text{I}(z,z_p)= 2 &\int \frac{\left(8\pi \alpha_s\right)^2}{s_{123}^2} \,C_F^2 \, \la P_{q\rightarrow g_1 g_2 q_3 }^{(ab)}\ra \sd\Phi_3\,\\
& \delta_\rho(1,2,3) \Theta(\theta_{23}<\theta_{13}) \Theta(\theta_{12}<\theta_{23})\\
& \delta(z(1-z_p)-z_2) \delta(z z_p-z_3).
\end{split}
\end{equation}
$\text{I}(z,z_p)$ has been written with a factor of two to account for both hemispheres and the angular integration extends only over the region relevant to the clustering of emissions 1 and 2. The simplified and correct calculations differ only in how soft the quark is allowed to be and in the limit $\zc \to 0$, the result after integration over $z$ and $z_p$ vanishes with $\zc$. This has been verified directly by numerical integration. Hence the correction term, given by Eq.~\eqref{eq:CfClustVanish}, can be neglected in our approximation.

The situation in the region $1-\zc<z<1$,  where we derive a correction to $\mathcal{F}^{\mathrm{fail}}(\rho,\epsilon)$, is somewhat more subtle. In the simplified version of the calculation, emission 1 is groomed away and the mass is set by emissions $2$ and $3$ which leads to a different constraint given by $\delta_\rho(2,3)$ on the angular integration compared to the correct version where all three partons are retained with $\delta_\rho(1,2,3)$, so one obtains a different result $\tilde{\text{I}}(z,z_p)$ given by:

\begin{equation}
\begin{split}
\tilde{\text{I}}(z,z_p)= 2&\int\frac{(8\pi \alpha_s)^2}{s_{123}^2} C_F^2 \la P_{q\rightarrow g_1 g_2 q_3 }^{(ab)}\ra\sd\Phi_3\,
\\
&\delta_\rho(2,3) \Theta(\theta_{23}<\theta_{13}) \Theta(\theta_{12}<\theta_{23})
\\
&\delta(z(1-z_p)-z_2) \delta(z z_p-z_3).
\end{split}
\end{equation}
Due to the fact that we are restricted to the angular region where the two gluons would be clustered, all angles are constrained to be small and we can use the triple-collinear limit for obtaining $\tilde{\text{I}}(z,z_p)$, ignoring the wide-angle modification required for the full calculation of $\mathcal{F}^{\mathrm{fail}}(\rho,\epsilon)$.

One key point here is that in the limit $z\to 1$, or equivalently $z_1 \to 0$, where there is a potential soft divergence, the condition $\delta_\rho (1,2,3)$ reduces to $\delta_\rho(2,3)$ so that the difference between $\tilde{\text{I}}(z,z_p)$ and $\text{I}(z,z_p)$ vanishes, leading to a finite result as already observed above. Moreover in the $\zc \to 0$  limit  we have that  $z \to 1$ over the full integration range so that one may simply replace $\tilde{\text{I}}(z,z_p)$ with $\text{I}(z,z_p)$ up to finite corrections suppressed by $\zc$. Doing so leads to
\begin{equation}\label{eq:CfClust1}
\begin{split}
\mathcal{F}^{\text{clust.}}_{C_F^2}  = \lim_{\zc\to 0}\int_{1-\zc}^1 &\left(\int_{\frac{\zc}{z}}^{\frac{1-\zc}{z}}\text{I}(z,z_p) \sd z_p
\right.\\
&\left.- \int_{\zc}^{1-\zc}\tilde{\text{I}}(z,z_p) \sd z_p \right) \sd z\\
&\hspace{-2.1cm}=\lim_{\zc\to 0}\int_{1-\zc}^1\sd z \int_{1-\zc}^{\frac{1-\zc}{z}}\text{I}(z,z_p)\sd z_p,
\end{split}
\end{equation}
where  in writing the third line we have replaced $\tilde{\text{I}}(z,z_p)$ with $\text{I}(z,z_p)$  and have exploited the fact that the difference between the lower limits of $z_p$ integration in the correct and subtracted term, corresponding to the region of a soft quark, leads only to terms power suppressed in $\zc$. We can numerically evaluate the integrals for a given $\zc$ value and on  decreasing $\zc$,  to  reduce the impact of power suppressed terms, we find the result converges to a constant. For our smallest value $\zc = 10^{-5}$, using the numerical method suave \cite{Hahn_2005}, we obtain
\begin{equation}
\mathcal{F}^{\text{clust.}}_{C_F^2} = \left(\frac{C_F \alpha_s}{2\pi}\right)^2 \left(4.246 \pm 0.002 \right).
\end{equation}

The  smallest value of $\zc$ was  chosen so that the error on the numerical integration was larger than the difference between the central values for the lowest and second lowest $\zc$ values.

The fact that the result for $\mathcal{F}^{\text{clust.}}_{C_F^2}$ tends to a constant as $\zc \to 0$ is related to the behaviour of the integrand in the soft limit for both emissions $z_p \to 1$ and $z  \to1$, and this allows us to also extract the result analytically. A series expansion of $\text{I}(z,z_p)$ around $z=1$ reveals a leading behaviour $\propto \frac{1}{(1-z)(1-z_p)}$, which derives from the soft limit of the matrix-element, and generates the full result in the limit $\zc \to 0$. It is straightforward to perform the integrals analytically to obtain:
\begin{equation}
\label{eq:cluster}
\mathcal{F}^{\text{clust.}}_{C_F^2} =\left(\frac{C_F \alpha_s}{2\pi} \right)^2  \frac{\pi}{12 \sqrt{3}}\left(3\psi^{(1)} \left ( \frac{1}{3} \right)- \psi^{(1)} \left (\frac{5}{6} \right) \right),
\end{equation}
where we have expressed the result in terms of the Polygamma function  $\psi^{(1)}(x)$. Note that one can further write
\begin{equation}
\begin{split}
\frac{\pi}{12 \sqrt{3}}\left(3\psi^{(1)} \left ( \frac{1}{3} \right)- \psi^{(1)} \left (\frac{5}{6} \right) \right) &=
\frac{4\pi}{3}\text{Cl}_2\left(\frac{\pi}{3}\right) \\
&=  4.25138 \cdots
\end{split}
\end{equation}
to obtain a compact result in terms of the Clausen function $\text{Cl}_2(x)$ \cite{wolfram}. This analytic result is consistent with the value obtained numerically for $\zc \to 10^{-5}$, keeping in mind that the latter includes power suppressed in $\zc$ terms varying as $\zc \ln^2\zc$. We also note that ignoring a region of phase space which only contributes a power of $\zc$ to $\mathcal{F}^{\text{clust}}$, the limits on the energy fraction integrals of Eq. \eqref{eq:CfClust1} can be re-written in terms of $z_1$ and $z_2$:
\begin{equation}
z_1<\zc , \qquad z_2<\zc , \qquad z_1+z_2>\zc.
\end{equation}
It is now apparent that clustering the two emissions together only leads to differences from our simplified treatment of the tagger where \textit{both} emissions would separately have failed the $\zc$ condition, but together lead to a cluster which passes the $\zc$ condition.

Our result for the  $\mathcal{F}^{\text{clust.}}_{C_F^2}$ turns out to be precisely the same as the result calculated previously, for the corresponding contribution to the non-cusp global soft anomalous dimension for the mMDT jet mass in SCET \cite{Frye:2016aiz,Bell:2018vaa}. While our starting point using the triple-collinear splitting functions goes beyond just the soft limit, the observation made above that the relevant limit for $\mathcal{F}^{\text{clust.}}_{C_F^2}$ is the limit when emissions $2$ and $3$ are additionally soft, explains the agreement with the soft limit calculations of  Refs.~\cite{Frye:2016aiz,Bell:2018vaa}. However it is worth stressing that our approach based on triple-collinear splitting functions remains valid beyond the soft limit and hence can also be used to compute the finite $\zc$ corrections that we have neglected in the present article.

\subsubsection{Virtual corrections and combined result}

Here we combine the results for double-real emission with the one-real one-virtual corrections to generate a finite result. The one-real--one-virtual terms are provided in appendix \ref{sec:r1v1}. We define the integral of the one-real--one-virtual term over $z$ as $V^{C_F^2}$, given by
\begin{equation}
V^{C_F^2}(\rho,\epsilon) = \int_{\zc}^{1-\zc} \sd{z} \, \mathcal{V}^{C_F^2}_{1,1}(\rho,z,\epsilon),
\end{equation}
with  $\mathcal{V}^{C_F^2}_{1,1}(\rho,z,\epsilon)$ given in Eq.~\eqref{eq:VirtualsCF}. Combining terms, we can write the gluon emission contribution as
\begin{equation}
\begin{split}
&\left(\rho \frac{d\Sigma_2}{d\rho}\right)^\text{gluon-emission} = \mathcal{F}^{\mathrm{opp.}}(\rho,\epsilon)
+\mathcal{F}^{\mathrm{pass}}(\rho,\epsilon)
\\
&\hspace{2.5cm}+\mathcal{F}^{\mathrm{fail}}(\rho,\epsilon)+\mathcal{F}^{\mathrm{clust.}}_{C_F^2}+ V^{C_F^2}(\rho,\epsilon).
\end{split}
\end{equation}
We find that after cancellation of all the singular contributions we are left with
\begin{equation}
\label{eq:emission}
\begin{split}
\left(\rho \frac{d\Sigma_2}{d\rho}\right)^\text{gluon-emission} &=
\left(\frac{C_F  \alpha_s}{2\pi} \right)^2 \bigg((3+4\ln\zc)^2\ln\rho \\
&-8 \ln ^3\zc-2 (3+16 \ln 2) \ln ^2\zc\\
&+(4-48 \ln2) \ln \zc \\
&-24 \zeta(3)+2 \pi ^2+\frac{7}{2}-18 \ln2 \\
&+\frac{4\pi}{3}\text{Cl}_2\left(\frac{\pi}{3}\right)+1.866 \pm 0.002\bigg),
\end{split}
\end{equation}
where the numerically quoted value $1.866 \pm 0.002$ represents the contribution to $\mathcal{F}^{\text{pass}}(\rho,\epsilon)$  arising from $H^{\text{fin.}}(z)$ (see Eqs.~\eqref{eq:zpass}, \eqref{eq:finitezpass}). It is then evident that the terms in Eq.~\eqref{eq:emission} that depend on $\ln \rho$ and $\ln \zc$ are  in precise agreement with those expected from the $C_F^2$ term in the expansion of the leading-log resummed result, i.e. Eq.~\eqref{eq:llexpansion}. In addition there is a constant contribution corresponding to an $\alpha_s^2 \ln  \rho$ NLL term  in $\Sigma_2(\rho)$. We shall analyse the constant contribution in more detail, after including another $C_F^2$ term coming from the gluon decay terms computed in the next sub-section.

\subsection{Gluon decay contributions}
Here we consider the contributions that are associated to the decay of a collinear gluon, emitted off the initiating quark, into a $q\bar{q}$ pair and a gluon pair associated with $C_F T_R n_f$ and $C_F C_A$ factors respectively.
For the case of the gluon decay to $q\bar{q}$ with a quark initiated jet i.e. a $q \to q \bar{q} q$ process, there is also an interference contribution from identical fermions in the final state with a colour factor $C_F (C_F-C_A)/2$, which contributes to the overall results for the $C_F^2$ and $C_F C_A$ channels. We shall first discuss this  piece and then turn to the $C_F T_R n_f$ and $C_F  C_A$ terms.

\subsubsection{$C_F \left(C_F-\frac{C_A}{2} \right)$ contribution}
The identical fermion contribution is simple to compute since it is finite both for the angular and energy integrals. The calculation can therefore be easily carried out numerically in four dimensions. The relevant splitting function is given in Eq.~\eqref{idensf} and we set $\epsilon \to 0$. Moreover due to the fact that the splitting function is regular in the energy fractions, the contribution from the region of integration where any parton has energy fraction  $z <\zc$ is suppressed with $\zc$. For this reason the clustering and grooming sequence does not matter as the result in the small $\zc$ limit comes from a configuration when all three partons contribute to the jet mass $\rho$  and have energy fraction $z>\zc$ i.e. the ungroomed limit. To obtain the leading term, which is a constant in the small $\zc$ limit, we set $\zc=0$ and numerically  perform the integral using our general rescaling method discussed in the appendix. We then have
\begin{equation}\label{eq:identical}
\begin{split}
\mathcal{F}^{\text{id}}&=\int \sd\Phi_3 \frac{\left(8 \pi \alpha_s \right)^2}{s_{123}^2} P^{(\text{id})}_{q\rightarrow q\bar{q}q} \,
\delta_\rho(1,2,3) = \\
&C_F\left(C_F-\frac{C_A}{2} \right)  \left( \frac{\alpha_s}{2\pi}\right)^2 \left (1.4386 \pm  0.0001 \right),
\end{split}
\end{equation}
where we performed the integral numerically with Suave and the result includes a factor of 2 to take account of both hemispheres as well an identical particle $1/2!$ phase-space symmetry factor. We believe that our result here coincides with an older calculation for the identical fermion contribution that enters initial state splittings, by Grazzini and de Florian, who obtained an analytical result
\begin{equation}
\label{eq:idanalyt}
\frac{13}{2}-\pi^2+4 \zeta(3)=1.43862 \cdots
\end{equation}
which they subtract to construct the relevant non-singlet contribution (see Eq.~(71) of Ref.~\cite{deFlorian:2001zd}). We shall return to this result, its analytical form and its combination with the $C_F^2$ and $C_F C_A$ results, when summarising our results.

\subsubsection{$C_F T_R n_f$ contribution}

\label{sec:nfpiece}
\begin{figure}[h]
\centering
 \includegraphics[width= 0.45\textwidth]{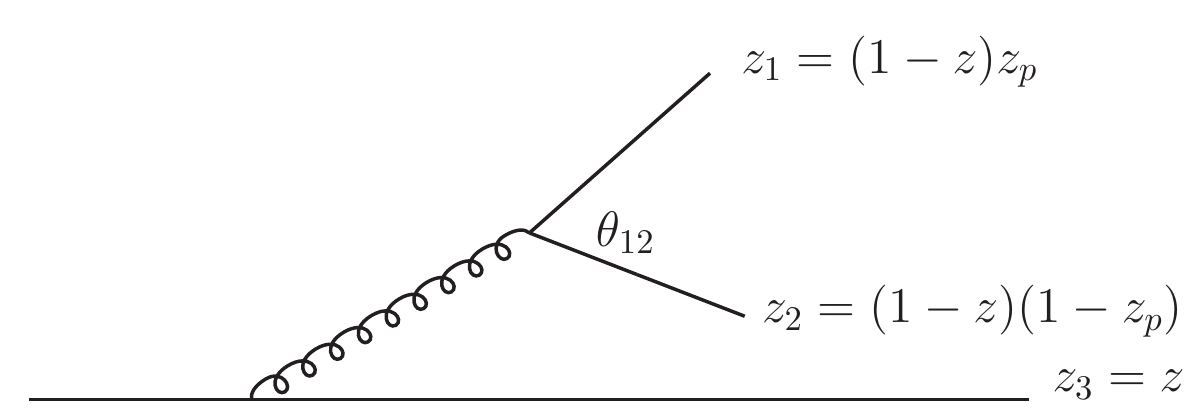}
 \caption{An illustration of the kinematic parametrisation in the $C_F T_R  n_f$ gluon decay channel.}
\label{fig:CFnf}
\end{figure}
Next we discuss the $C_F T_R n_f$ term again related to gluon decay to a $q\bar{q}$ pair. The relevant diagram with our parameterisation of the kinematics is shown in Figure \ref{fig:CFnf}. One encounters a collinear singularity in the squared matrix element as $\theta_{12} \to 0$, since the  $g \to q\bar{q}$ splitting is regular in the infrared, which leads to a $1/\epsilon$ pole for the jet mass distribution. We also expect that an analysis of the gluon decay contributions should lead to the emergence of the correct argument of the running coupling i.e. $k_t$ in the soft limit, and the factor associated to the physical CMW scheme.

As we did for the $C_F^2$ piece, it proves convenient to break the calculation into two pieces : a first piece that simplifies the action of the tagger and contains the divergent pole structure and a correction term which only has support in non-singular regions, leading to a finite result which can be computed numerically in four dimensions. We will also account for virtual corrections which cancel the pole in the real emission piece, leaving a finite result. The divergence occurs as
$\theta_{12} \to 0$ which is the region where the $q$ and $\bar{q}$ are clustered first in the C/A algorithm, and then the $q\bar{q}$ pair is clustered to the parton $3$. We shall therefore first carry out the calculation always taking the $q$ and $\bar{q}$ to be clustered together which mistreats regions where emissions $1$ or $2$ could be first clustered with $3$. These regions will be subject to our finite clustering corrections.

For the contribution clustering the $q$ and $\bar{q}$ from the gluon decay, on declustering we produce two branches consisting  of the quark (or anti-quark labeled) $3$  and the clustered fermion pair or equivalently the massive parent gluon.  If the quark  labeled $3$ is soft and fails the $\zc$ then we obtain only terms power suppressed in  $\zc$. If the massive gluon branch fails the clustering we obtain a massless hemisphere. Thus we have a situation where in order to obtain a finite result in the $\zc \to 0$ limit, both branches pass the $\zc$ condition so that all three partons are retained and contribute to the jet mass. The condition $\rho \ll \zc$ once again implies that all three partons are collinear and we can apply the triple-collinear splitting function and phase-space. In order to use our integration strategy based on rescaled angular variables, we consider two regions $\theta_{12}^2<\theta_{13}^2$ and $\theta_{12}^2>\theta_{13}^2$, with the first mentioned region contributing the divergence. We denote the respective contributions to $\rho \frac{d\Sigma_2}{d\rho}$ by $\mathcal{F}^{\theta_{12}<\theta_{13}}(\rho)$ and $\mathcal{F}^{\theta_{12}>\theta_{13}}(\rho)$  where explicitly we have for the former,
\begin{equation}
\begin{split}
\mathcal{F}^{\theta_{12}<\theta_{13}}(\rho,\epsilon) = 2 S_\epsilon^{-2} &\left(\frac{Q}{2}\right)^{4\epsilon}
\int \sd \Phi_3 \frac{\left(8 \pi \alpha_s \right)^2}{s_{123}^2}  \la P_{\bar q_1^\prime q^\prime_2 q_3 }\ra
\\
&\delta_\rho(1,2,3) \Theta_{\zc}(12|3) \Theta(\theta_{12}<\theta_{13}),
\end{split}
\end{equation}
where a sum over flavours leading  to a factor $n_f$ is left implicit on the RHS of the above equation. The notation $\Theta_{\zc}(12|3)$ denotes that each of the two branches passes the $\zc$ condition equivalent to the constraint on the parent gluon energy $ 1-\zc> z >\zc$, the result is written in terms of the renormalised $\overline{\text{MS}}$ coupling and a factor of two accounts for both hemispheres.

In terms of the rescaled angular variable $y=\theta_{12}^2/\theta_{13}^2$ and the parent energy fraction $z$, we obtain a result of the form (where $G(y,z,\epsilon)$ is  regular as $y \to 0$)
\begin{equation}
\label{eq:nfpole}
\begin{split}
\mathcal{F}^{\theta_{12}<\theta_{13}}(\rho,\epsilon)&=\rho^{-2\epsilon}\int_{\zc}^{1-\zc}\sd z\int_0^1\frac{\sd y}{y^{1+\epsilon}}G(y,z,\epsilon) \\
& -\frac{\rho^{-2\epsilon}}{\epsilon}\int_{\zc}^{1-\zc}\sd z\ G(0,z,\epsilon)\\
&+\int_{\zc}^{1-\zc}\sd z\int_0^1 \frac{dy}{y}\left(G(y,z,0)-G(0,z,0) \right).
\end{split}
\end{equation}
In writing the above we first isolated the singular contribution by taking only the leading term in the expansion of $G(y,z,\epsilon)$ around $y=0$, and integrated over $y$ to obtain the pole in $\epsilon$  in the first term above, while the second term is finite by construction since $\left(G(y,z,0)-G(0,z,0) \right)$ vanishes as $y \to 0$. The finite term can be computed in four dimensions, so we set  $\epsilon=0$.
However while the term involving $\left(G(y,z,0)-G(0,z,0) \right)$ is finite, it has a leading behaviour in the $z \to 1$ limit proportional to $\frac{1}{1-z}$, which can be extracted through a series expansion about $z=1$. The limit $z \to 1$ corresponds to a soft parent gluon and gives rise to $\ln \zc$ terms in the result, which build up the constant  $K$ which  relates the coupling in the $\overline{{{\text{MS}}}}$ scheme to that in the CMW scheme.  After separating the $1/(1-z)$ term which can be handled analytically, we integrate the remainder of the finite contribution numerically.

We then have
\begin{equation}
\begin{split}
&\mathcal{F}^{\theta_{12}<\theta_{13}}(\rho,\epsilon)= C_F T_R n_f \left( \frac{\alpha_s}{2\pi} \right)^2 \times
\\
&\int_{\zc}^{1-\zc} \left(\frac{H_1^{\text{coll.}}(z,\rho,\epsilon)}{\epsilon}
 +H_1^{\text{fin.-soft}}(z)+H_1^{\text{fin.}}(z) \right)\ \sd z\ ,
\end{split}
\end{equation}
where
\begin{equation}
\begin{split}
&H_1^{\text{coll.}}(z,\rho,\epsilon)= \\
&\rho^{-2\epsilon}(1-z)^{-2\epsilon} p_{qq}(z,\epsilon) \left(-\frac{4}{3}-\frac{46}{9}\epsilon+\mathcal{O}(\epsilon^2)\right),\\
&H_1^{\text{fin.-soft}}(z)=\frac{4}{9}\frac{1}{1-z}.
\end{split}
\end{equation}
$H_1^{\text{fin.-soft}}(z)$ is the soft parent finite contribution and the finite remainder $H^{\text{fin.}}(z)$ tends to a constant as $\zc \to 0$. Using Mathematica's NIntegrate \cite{Mathematica} with $\zc=0$ we obtain
\begin{equation}
\int_0^1H_1^{\text{fin.}}(z) \sd z=-1.479\pm0.001 \ .
\end{equation}

Next we need the contribution from the angular region $\theta_{13}<\theta_{12}$ which does not contain any poles. However, as discussed above for  $\theta_{12}<\theta_{13}$, there is again a soft enhancement as $z\to1$ giving rise to a $\ln \zc$ term related to the CMW constant $K$. Explicitly we have:
\begin{equation}
\begin{split}
&\mathcal{F}^{\theta_{13}<\theta_{12}}(\rho)= \\
&2  \int \sd \Phi_3 \frac{\left(8\pi \alpha_s\right)^2}{s_{123}^2}   \la P_{\bar q_1^\prime q^\prime_2 q_3 }\ra \delta_\rho(1,2,3) \Theta_{\zc}(12|3) \Theta(\theta_{13}<\theta_{12}),\\
&=C_F T_R n_f \left( \frac{\alpha_s}{2\pi} \right)^2 \int_{\zc}^{1-\zc} \left(H_2^{\text{fin.-soft}}(z)+H_2^{\text{fin.}}(z) \right)\ \sd z\,
\end{split}
\end{equation}
which has no collinear pole i.e. $H_2^{\text{coll.}}(z)=0$ and where
\begin{equation}
\begin{split}
H_2^{\text{fin.-soft}}(z)&=\frac{16}{3}\frac{1}{1-z}, \\
\int_0^1H_2^{\text{fin.}}(z) \sd z &=  -6.242\pm0.006.
\end{split}
\end{equation}
where, as before, the final term is evaluated numerically with $\zc=0$ to remove power corrections.

A final ingredient for generating the result is the $T_R n_f$ piece of the one-real--one-virtual contribution. This is reported in \ref{sec:r1v1} (see Eq.~\eqref{eq:nfvirtual}) and a finite result is obtained on combining the virtual contribution with $\mathcal{F}^{\theta_{13}>\theta_{12}}(\rho)$. Defining
\begin{equation}
V^{C_F T_R n_f}\left(\rho,\epsilon \right) = \int_{\zc}^{1-\zc} \sd z  \mathcal{V}_{1,1}^{C_FT_R n_f}\left(\rho,z,\epsilon  \right)
\end{equation}
and
\begin{equation}
\mathcal{F}^{\mathrm{tot.}}(\rho,\epsilon)=\mathcal{F}^{\theta_{12}<\theta_{13}}(\rho,\epsilon) +\mathcal{F}^{\theta_{13}<\theta_{12}}(\rho),
\end{equation}
we can combine the results to obtain
\begin{eqnarray}
\label{eq:nftot}
&&\mathcal{F}^{\mathrm{tot.}}(\rho) + V^{C_FT_R n_f}\left(\rho,\epsilon \right) = \nonumber\\
&&C_F T_R n_f \left (\frac{\alpha_s}{2\pi} \right)^2 \left( \int_{\zc}^{1-\zc} \sd z \, G(z,\rho) -7.721 \pm 0.007\right),
\nonumber\\
\end{eqnarray}
where
\begin{equation}
\begin{split}
G(z,\rho) &= \frac{4}{3} p_{qq}(z) \ln(\rho(1-z)) -\frac{20}{9} p_{qq}(z)\\
&-\frac{4}{3} p_{qq}(z) \ln z+\frac{26}{9}(1+z).
\end{split}
\end{equation}
It is worth making some remarks on the  form of $G(z,\rho)$. Firstly we note a piece corresponding to the $n_f$ term in $-2  \times p_{qq} b_0 \ln \left(\rho(1-z)\right)$, where $b_0 =\frac{11}{6} C_A-\frac{2}{3} T_R n_f $ is the first perturbative coefficient of the QCD $\beta$ function.  This term produces an LL contribution from the dependence on $\ln \rho$. In the soft limit, i.e. $z \to 1$ and $p_{qq}(z)  \to \frac{2}{1-z}$ it can be absorbed into the leading-order result by correcting the scale of the coupling $\alpha_s \left(\frac{Q^2}{4}\right) \to \alpha_s \left(\frac{Q^2}{4} \rho (1-z)\right) =\alpha_s(k_t^2)$, consistent with the LL formula $\eqref{eq:llresult}$. The term $-\frac{20}{9} p_{qq}(z)$ corresponds to the appearance of the $n_f$ piece of the CMW constant $K$, as also anticipated in the LL result.
The remaining terms produce a constant in the small $\zc$ limit, on integration over $z$, which can be combined with the constants we obtained numerically. However before doing so we shall evaluate the correction term due to the proper use of the C/A clustering sequence within the mMDT.

Turning to the clustering correction we first note that in the angular region where  $\theta_{12}$ is smallest our treatment of the tagger, working as if partons 1 and 2 are always clustered, needs no correction. The regions where a correction is needed are when $\theta_{13}$ is smallest and when $\theta_{23}$ is smallest which is identical due to the symmetry under $1 \leftrightarrow 2$. These  regions contain no divergences and hence the calculation of the correction term, i.e. the difference between the correct and simplified calculations, can be performed numerically in four dimensions. We shall also work in the limit $\zc \to 0$ to eliminate power-suppressed terms in $\zc$, explicitly take the case that $\theta_{13}$ is the smallest angle and double the result to account for $\theta_{23}$ being smallest. We can then write
\begin{equation}\label{eq:clustnf}
\begin{split}
\mathcal{F}^{\text{clust.}}_{C_F T_R n_f} = \lim_{\zc\to 0} 4  &\int \frac{\left(8\pi \alpha_s\right)^2}{s_{123}^2} \, \la P_{\bar q_1^\prime q^\prime_2 q_3 }\ra \sd\Phi_3 \, \Theta(\theta_{23}>\theta_{13})  \\
&\Theta(\theta_{12}>\theta_{13}) \bigg( \delta_\rho(1,2,3) \Theta_{\zc}(2|13) \\
&\hspace{2cm}+ \delta_\rho(1,3) \Theta_{\zc}(1|3) \\
&\hspace{2cm}- \delta_\rho(1,2,3) \Theta_{\zc}(3|12) \bigg).
\end{split}
\end{equation}
The above equation reflects that when $\theta_{13}$ is the smallest angle, there are two configurations that yield a massive hemisphere: when parton 2 passes the $\zc$ condition and all three partons are retained (the first term on the second line), and when parton 2 fails the $\zc$ condition but partons 1 and 3 pass (the second term on the second line). The clustering correction may then be expressed as
\begin{equation}\label{eq:nfClust}
\begin{split}
\mathcal{F}^{\text{clust.}}_{C_F T_R n_f}=\lim_{\zc\to 0} \int_0^1 &\left(\text{I}_{n_f}(z,z_p)\left(\Theta_{\zc}(13|2)-\Theta_{\zc}(12|3)\right)
\right.\\
&\left.+\tilde{\text{I}}_{n_f}(z,z_p)\Theta_{\zc}(1|3) \right) \sd z \sd z_p\ ,
\end{split}
\end{equation}
where the subtraction term corresponds to removal of the simplified contribution where 1 and  2 were taken to be clustered first and our usual notation applies where $ \Theta_{\zc}(a|b)$ denotes the two branches $a$ and $b$ that pass the $\zc$ condition in each case.\footnote{This implies also the condition that partons not included in these branches fail the  $\zc$ e.g. for parton $2$ in $\Theta_{\zc}(1|3)$.} The integrals $\text{I}_{n_f}(z,z_p)$ and $\tilde{\text{I}}_{n_f}(z,z_p)$ arise from the angular integration for the case when all three partons contribute to the jet mass and when only two partons contribute respectively.\footnote{In the case when a parton is soft enough to be groomed away it is not constrained by the jet mass and can in principle be at a large angle. For correlated emission such configurations, where one of the gluon decay offspring is at a large angle to the other, are dynamically suppressed and only contribute at the level of power corrections in $\rho$. This implies that the integrals converge within the triple-collinear region which does not need to be modified, which we have also verified numerically.} Explicitly we have
\begin{equation}\label{eq:Idef}
\begin{split}
\text{I}_{n_f}(z,z_p)= 4\int \frac{(8\pi \alpha_s)^2}{s_{123}^2} &\la P_{q\rightarrow \bar{q}'_1 q'_2 q_3 }\ra \sd \Phi_3\  \delta_\rho(1,2,3)\\
&\Theta(\theta_{13}<\theta_{23}) \Theta(\theta_{13}<\theta_{12}) \\
&\times\delta(z(1-z_p)-z_2) \delta(z z_p-z_3) \ \ ,
\end{split}
\end{equation}

and

\begin{equation}\label{eq:ITildeDef}
\begin{split}
\tilde{\text{I}}_{n_f}(z,z_p)= 4\int\frac{(8\pi \alpha_s)^2}{s_{123}^2} &\la P_{q\rightarrow \bar{q}'_1 q'_2 q_3 }\ra \sd\Phi_3\  \delta_\rho(1,3)
\\
&\Theta(\theta_{13}<\theta_{23}) \Theta(\theta_{13}<\theta_{12}) \\
&\times \delta(z(1-z_p)-z_2) \delta(z z_p-z_3) \ \ ,
\end{split}
\end{equation}
where a factor of $4$ accounts for both hemispheres and the case when $\theta_{23}$ is the smallest angle.

We can evaluate the integrals numerically by choosing a small $\zc$ to suppress power corrections and find that the result tends to a  $\zc$ independent constant on decreasing $\zc$. On evaluation of the integrals numerically with Suave we obtain, for $\zc=10^{-5}$ as for the $C_F^2$ clustering piece, the result
\begin{equation}
\label{eq:nfclust}
\mathcal{F}^{\text{clust.}}_{C_F T_R n_f} = C_F T_R n_f \left(\frac{\alpha_s}{2\pi}\right)^2 \left(-1.754\pm  0.002 \right).
\end{equation}

Further insight into the nature of the clustering correction reported above can be obtained via similar considerations to those for the $C_F^2$ clustering correction. Firstly one notes that in the limit parton $2$ goes soft and fails the $\zc$ condition the angular integral $\tilde{\text{I}}_{n_f}(z,z_p)$ may be replaced by $\text{I}_{n_f}(z,z_p)$ up to terms that vanish with $\zc$. This lets us combine the constraints on the $z$ integrals and, again, with neglect of power corrections in $\zc$ one obtains the conditions  on $z_1$ and $z_2$:
\begin{equation}
z_1+z_2>\zc \ , \qquad z_1<\zc \ , \qquad z_2<\zc.
\end{equation}
These conditions are the same as for the $C_F^2$ case, however here the clustering correction enters with a negative sign while a positive correction was noted for the  $C_F^2$ term. The reason for this is that our simplified treatment in the  $C_F^2$ channel amounted to discarding two emissions that individually failed the $\zc$, thereby {\it{excluding}} the contribution where they pass the $\zc$ when correctly treated as a cluster since $z_1+z_2>\zc$. Here on the other hand, our simplified picture {\it{includes}} configurations where incorrectly treating emissions as a cluster they pass  the $\zc$ condition, while in the correct treatment of clustering where the gluon decay products are not clustered the emissions each fail the  $\zc$ leading to a massless hemisphere.

Furthermore, we note once again that our clustering corrections originate in the soft region, albeit still also within the triple-collinear regime and that our calculations can be extended to include finite $\zc$ terms. The numerical value we obtain for the clustering piece,  $\mathcal{F}^{\text{clust.}}_{C_F T_R n_f}$, is once again in agreement within errors (and potential $\zc \ln^2 \zc$ terms) to that previously obtained for the $n_f$ part of the clustering term for the non-cusp global soft anomalous dimension for mMDT in the SCET framework \cite{Frye:2016aiz,Bell:2018vaa}.

Finally we quote our overall result for the $C_F T_R n_f$ channel, combining the different contributions i.e. performing the integral over $z$ in Eq.~\eqref{eq:nftot} and adding in the clustering correction:
\begin{equation}
\label{eq:nfres}
\begin{split}
\rho \frac{d\Sigma^{C_F T_Rn_f}_2}{d\rho} &=  C_F T_R n_f \left( \frac{\alpha_s}{2\pi} \right)^2 \times \\
&\bigg(-\frac{2}{3}(3+4\ln\zc)\ln\rho
-\frac{4}{3}\ln^2\zc+\frac{40}{9}\ln\zc \\
&+\frac{4  \pi^2}{9}+\frac{25}{3}-7.721 \pm 0.007 -1.754 \pm 0.002 \bigg),
\end{split}
\end{equation}
where the $-1.754$ is the clustering correction computed above, and the other terms are the full result for our simplified treatment of the tagger. We note that  the $\ln\rho$ and $\ln \zc$ dependent terms in Eq.~\eqref{eq:nfres} are in exact agreement with expectations from the expansion of the leading logarithmic result \eqref{eq:llexpansion}. We shall shed further light on the constant term $\frac{4  \pi^2}{9}+\frac{25}{3}-7.721\pm 0.007$ following an analytic calculation in the next subsection.

\subsubsection{Analytic calculation using web variables}

Our aim is to compute the simplified piece of the differential distribution, i.e. $\mathcal{F}^{\text{tot.}}$, which neglects the proper C/A clustering in specific angular regions, using the web variables given in~\ref{app:web}. In particular, this allows for a complete analytic extraction of the constant in eq.~\eqref{eq:nftot}. In addition, the analytic computation offers insight into the structure of the answer for the $C_F C_A$ channel to be presented in section~\ref{cfca}, which shall become evident when we discuss the final results.

First, we recall
\begin{equation}\label{eq:maindist}
\begin{split}
\mathcal{F}^{\text{tot.}}(\rho)&=2 S_\epsilon^{-2}\left( \frac{Q}{2}\right)^{4\epsilon}\times\\
&\int \sd \Phi_3 \frac{\left(8 \pi \alpha_s \right)^2}{s_{123}^2}  \la P_{{\bar q}^\prime_1
q^\prime_2 q_3} \ra \delta_\rho(1,2,3)\, \Theta_{\zc}(12|3)  \ \ .
\end{split}
\end{equation}
With few steps one can express the triple-collinear splitting function in terms of the web variables. In particular, we have
\begin{equation}
\begin{split}
t_{12,3} &= - 2 \frac{z}{1-z} z_p (1-z_p) \times\\
&\left[\frac{1-2z_p}{z_p(1-z_p)} s_{12} + \frac{2 k_t \sqrt{s_{12}}}{\sqrt{z_p(1-z_p)}} \cos\theta\right] - (1-2z_p) s_{12} \ \ ,
\end{split}
\end{equation}
where $\theta \in [0,\pi]$ is the polar angle in the $(2-2\epsilon)$-dimensional transverse plane, i.e. the opening angle between $q_\perp$ and $k_\perp$. In addition, the jet mass reads
\begin{align}\label{jetmass}
s_{123} = s_{12} + \frac{z}{1-z} s_{12} + \frac{z}{1-z} k_t^2 \ \ ,
\end{align}
and here we observe the first simplification, namely that the jet mass does not depend explicitly either on $z_p$ or $\theta$, consequently, the latter variables can be integrated from the outset\footnote{A caveat exists in the presence of soft divergences, e.g. for the $C_F C_A$ channel. In this case, one has to re-express the phase space in terms of the angle $\theta_{12}$ instead of the invariant mass $s_{12}$. This is because $s_{12} \propto z_p(1-z_p) \theta^2_{12}$, which affects the $\epsilon$-dependent integration measure over $z_p$.}. One can partition the computation initially into two pieces, the first contains the collinear divergence while the second is collinear-regular.

Let us start with the divergent piece which, after angular averaging over $\theta$, reads
\begin{align}
\nonumber
 \mathcal{F}_{\text{pole}}^{\text{tot.}}(\rho) &=  C_F T_R n_f \left( \frac{\alpha_s}{2\pi} \right)^2 4^{-2\epsilon}\times
\\\nonumber
 &\int_{\zc}^{1-\zc} \,   \frac{ dz z^{1-2\epsilon}}{1-z} \frac{dk_t^2}{k_t^{2\epsilon}} \frac{ds_{12}}{(s_{12})^{1+\epsilon}} \frac{dz_p}{ (z_p(1-z_p))^{\epsilon}}\\\nonumber
& \times \left[ - 8 \frac{z}{1-z} \frac{z_p (1-z_p)}{1-\epsilon} + \frac{4z}{1-z}\right. \\\nonumber
& \left. + (1-z) (1-2z_p)^2 + (1-2\epsilon)(1-z) \vphantom{\frac{z}{1-z}}    \right] \\
 &\times \delta\left(\frac{s_{12}}{1-z} + \frac{z k_t^2}{1-z} - \frac{\rho}{4} \right) \ \ ,
\end{align}
where (and in what follows below) we rescaled all dimensionful quantities by $Q^2$. In the above equation we can observe another simplification, namely the double-soft limit, i.e. $z \to 1$, is quite transparent. One can rewrite the integral, making sure to isolate the double-soft behaviour, as follows
\begin{align}\label{eq:cfnfdsoft}
\nonumber
 \mathcal{F}_{\text{pole}}^{\text{tot.}}(\rho) &= C_F T_R n_f \left( \frac{\alpha_s}{2\pi} \right)^2 4^{-2\epsilon}\times
 \\\nonumber
 &\int_{\zc}^{1-\zc}  \,   \frac{ dz z^{1-2\epsilon}}{1-z} \frac{dk_t^2}{k_t^{2\epsilon}} \frac{ds_{12}}{(s_{12})^{1+\epsilon}} \frac{dz_p}{ (z_p(1-z_p))^{\epsilon}}\\\nonumber
& \times \left[ 4 \frac{p_{\text{qg}}(z_p,\epsilon)}{1-z}- 4 p_{\text{qg}}(z_p,\epsilon) + (1-z) (1-2z_p)^2 \right.\\
&\left.+ (1-2\epsilon)(1-z) \vphantom{\frac{p_{\text{qg}}(z_p,\epsilon)}{1-z}} \right]
\delta\left(\frac{s_{12}}{1-z} + \frac{z k_t^2}{1-z} -\frac{\rho}{4} \right) \ \ ,
\end{align}
where the $g \to q\bar{q}$ $\sd$-dimensional splitting function has been identified
\begin{align}
p_{\text{qg}}(z_p,\epsilon) = 1 - 2 \frac{z_p(1-z_p)}{1-\epsilon} \ \ .
\end{align}
Now the double-soft piece in eq.~\eqref{eq:cfnfdsoft} can be integrated directly over $z_p$ to find the CMW coupling. Finally we add the contribution in $ \mathcal{F}^{\text{tot.}}$ which do not exhibit a collinear pole, i.e.
\begin{align}
\mathcal{F}^{\text{tot.}}_{\text{reg.}}(\rho) =   - C_F T_R n_f \left( \frac{\alpha_s}{2\pi} \right)^2\int_{\zc}^{1-\zc}  dz \, \frac43 (1-z) \ \ ,
\end{align}
where we performed the integrals over $k_t$, $s_{12}$ and $z_p$. The final result is obtained after adding in the virtual corrections, and we get
\begin{equation}\label{eq:nffinal}
\begin{split}
\mathcal{F}^{\text{tot.}}(\rho) &= 2 \,C_F T_R n_f \left( \frac{\alpha_s}{2\pi} \right)^2 \times\\
\int_{\zc}^{1-\zc} dz &\bigg[ \frac23 p_{\text{qq}}(z) \ln(\rho(1-z)) - \frac{10}{9} \frac{2}{1-z}\\
& + \frac{10}{9} (1+z) - \frac23 (1-z) \bigg] \ \ .
\end{split}
\end{equation}
An important feature of the analytic result is the disappearance of any $\ln(z)$ terms as they fully cancel between real and virtual corrections. These structures, if present, would have led to $\zeta(2)$ when integrated against $p_{\text{qq}}(z)$. We can now include the clustering correction and integrate eq.~\eqref{eq:nffinal} over $z$ (dropping power corrections in $z_{\text{cut}}$) to find
\begin{equation}\label{eq:nffinal2}
\begin{split}
\rho \frac{d\Sigma^{C_F T_Rn_f}_2}{d\rho} &=  C_F T_R n_f \left( \frac{\alpha_s}{2\pi} \right)^2 \bigg(-\frac{2}{3}(3+4\ln\zc)\ln\rho
\\
&-\frac{4}{3}\ln^2\zc +\frac{40}{9}\ln\zc + 5 - 1.754 \pm 0.002 \bigg)  ,
\end{split}
\end{equation}
which is fully consistent with eq.~\eqref{eq:nfres}.

\subsubsection{$C_F C_A$ contribution from $q \to  q gg$\label{cfca}}

The same kinematic variables apply as in Fig.~\ref{fig:CFnf} for the gluon decay to $q\bar{q}$. One of the key differences with the $n_f$ piece is now the presence of soft divergences as $z_p \to 0$ and $z_p \to 1$.\footnote{In order to avoid considering both limits one can simply take the region $z_p < 1-z_p$, where the divergence only comes from $z_p \to 0$, and double the result exploiting the symmetry between the gluons.}

We can organise the calculation in precisely the same way as for the $n_f$ piece by first computing a simplified term where for applying the grooming the offspring gluons are always treated as a cluster equivalent to the parent gluon. We then correct for the proper C/A clustering so that, as before, our correction term is finite and calculable in four dimensions. Again as done before for the $n_f$ piece, we can further  divide the simplified calculation into two pieces where $\theta_{12} < \theta_{13}$ and vice-versa. The region with  $\theta_{12} < \theta_{13}$ contains all the divergences,  resulting in $\frac{1}{\epsilon^2}$ and $\frac{1}{\epsilon}$ poles. The region $\theta_{13} > \theta_{12}$ gives only a finite contribution in spite of the presence of soft divergences in the $g  \to gg$ splitting, as a consequence of the angular ordering property of soft radiation.
\begin{equation}
\begin{split}
\mathcal{F}_{C_F C_A}^{\theta_{12}<\theta_{13}}(\rho,\epsilon) &= 2 S_\epsilon ^{-2} \left(\frac{Q}{2}\right)^{4\epsilon}
\int\sd \Phi_3 \frac{\left(8 \pi \alpha_s \right)^2}{s_{123}^2}  \la P^{(\text{nab})}_{q\rightarrow g_1 g_2 q_3 }\ra
\\
& \times \delta_\rho(1,2,3) \Theta_{\zc}(12|3) \Theta(\theta_{12}<\theta_{13}) \ ,
\end{split}
\end{equation}
which is written as before in terms of the renormalised $\overline{\mathrm{MS}}$ coupling with $\mu_R =Q/2$, and a factor of 2 to account for both hemispheres. Further analysis using our general integration method, outlined in the appendix, gives
\begin{equation}\label{eq:CaRealPoles}
\begin{split}
&\mathcal{F}_{C_F C_A}^{\theta_{12}<\theta_{13}}(\rho,\epsilon)=C_F C_A \left(\frac{\alpha_s}{2\pi} \right)^2\times\\
&\int_{\zc}^{1-\zc} \left(\frac{H^{\text{soft-coll.}}(z,\rho,\epsilon)}{\epsilon^2}
 +\frac{H^{\text{coll.}}(z,\rho,\epsilon)}{\epsilon}+H^{\text{finite}}(z)  \right) \sd z\ ,
\end{split}
\end{equation}
where
\begin{equation}
\begin{split}
H^{\text{soft-coll.}}(z,\rho,\epsilon)&= \rho^{-2\epsilon}(1-z)^{-2\epsilon}\times
\\
&p_{qq}(z,\epsilon) \left(2-\frac{\pi^{2}}{3}\epsilon^2 +\mathcal{O}(\epsilon^3)\right), \\
H^{\text{coll.}}(z,\rho,\epsilon)&= \rho^{-2\epsilon}(1-z)^{-2\epsilon}\times
\\
&p_{qq}(z,\epsilon) \left(\frac{11}{3}+\frac{134}{9}\epsilon-\frac{4\pi^2}{3}\epsilon +\mathcal{O}(\epsilon^2)\right).
\end{split}
\end{equation}
As in section \ref{sec:nfpiece} the finite term $H^{\text{finite}}(z)$ is enhanced in the limit of a soft parent, $z \to 1$, and produces $\ln \zc$ terms, which we wish to separate since they relate to the CMW scheme. As before we use a series expansion about $z=1$, to make the decomposition
\begin{equation}
H^{\text{finite}}(z)=\frac{c}{1-z} + f(z),
\end{equation}
where $f(z)$ is finite as $z \to 1$. The constant, $c$ is evaluated numerically (on integrating over the angular variables and $z_p$), as is the integral over $f(z)$ and, again using NIntegrate with $\zc=0$, we obtain:
\begin{equation}
\int_{\zc}^{1-\zc}H^{\text{finite}}(z)=(2.4361\pm0.0002)\ln\zc -0.117\pm0.001.
\end{equation}
In the region $\theta_{13}<\theta_{12}$ there are no poles in $\epsilon$ and we can perform the calculation setting $\epsilon \to 0$, so we have
\begin{equation}
\begin{split}
\mathcal{F}^{\theta_{13}<\theta_{12}}_{C_F  C_A} (\rho)=\int \sd &\Phi_3 \frac{\left(8 \pi \alpha_s \right)^2}{s_{123}^2}
\la P_{q\rightarrow g_1 g_2 q_3 }\ra \\
&\delta_\rho(1,2,3) \Theta_{\zc}(12|3) \Theta(\theta_{13}<\theta_{12}).
\end{split}
\end{equation}
Again separating the integrand into pieces which diverge as $z\to1$ and those which do not, we obtain, after numerical integration with $\zc=0$,
\begin{equation}
\begin{split}
\mathcal{F}_{C_F C_A}^{\theta_{13}<\theta_{12}}(\rho)=C_F C_A \left(\frac{\alpha_s}{2\pi} \right)^2
&\left[(5.8730\pm0.0006)\ln\zc \right.\\
&\left.+(6.795\pm0.006)\right].
\end{split}
\end{equation}

Finally we account for the clustering corrections.  This is done as for the $n_f$ piece in Eq.~\eqref{eq:clustnf} and the result is finite as soft divergences cancel in the combination of the correct and simplified treatments. The result can be computed numerically in four dimensions. Using Suave with $\zc=10^{-5}$ we obtain:
\begin{equation}\label{eq:CAclustfinal}
\mathcal{F}^{\text{clust.}}_{C_F C_A}=C_F C_A \left(\frac{\alpha_s}{2\pi} \right)^2\left(-1.161\pm0.001\right).
\end{equation}

The same comments apply to the origin of the clustering correction here as for the $n_f$ piece, namely it originates from incorrectly allowing, in the simplified result, the gluon pair to pass the clustering due to the fact that  the parent passes the $\zc$. The correct tagging procedure would be applied to the individual gluons instead, which fail the clustering leading to a massless jet and a nil contribution. The result again agrees with previous calculations of the clustering piece in the mMDT SCET non-cusp soft anomalous dimension \cite{Frye:2016aiz,Bell:2018vaa}  to within errors and potential $\zc \ln^2 \zc$ corrections.

Finally we combine all pieces and include the one-real--one-virtual correction $V^{C_F C_A}\left(\rho,\epsilon \right) = \int_{\zc}^{1-\zc} \sd z  \mathcal{V}_{1,1}^{C_F C_A}\left(\rho,z,\epsilon  \right) $ defined in the appendix to obtain the result
\begin{equation}\label{eq:CAfinal}
\begin{split}
&\rho \frac{d\Sigma_2}{d\rho}^{q \to qgg, \mathrm{nab.}} = \\
&C_F C_A \left(\frac{\alpha_s}{2\pi}\right)^{2}
\bigg[\frac{11}{6}\left( 3+4\ln\zc\right)\ln\rho
+\frac{11 \ln^2\zc}{3} \\
&\hspace{2cm}+\bigg(\frac{4}{3}\pi^{2} -\frac{268}{9} +(8.3091\pm0.0006)\bigg)\ln\zc \\
&\hspace{2cm}+16 \zeta (3)-\frac{11 \pi ^2}{9}-\frac{121}{6}\\
&\hspace{2cm}+(6.678\pm0.006)-(1.161\pm0.001)\bigg] \ \ ,
\end{split}
\end{equation}
 where the labelling $q \to qgg, \mathrm{nab.}$ indicates the non-abelian contribution to the $q  \to qgg$ process  and where we have separately written the numerically computed clustering contribution  and the numerically computed part of the simplified calculation.
 The result above is in agreement with our expectations from the expansion of the leading-logarithmic resummed result  Eq.~\eqref{eq:llexpansion} for the terms involving $\ln \rho$ and $\ln \zc$ since the numerical value $8.3091 \pm 0.0006$ is in good agreement with $\frac{134}{9}-\frac{2\pi^2}{3}$, signalling again that the $\ln \zc$ term is associated with the CMW scheme,   while the $\ln  \rho$ and $\ln^2 \zc$ terms are  associated to the argument of the running coupling i.e. $k_t$  in  the soft limit. This leaves us to comment on the constant term, other than the clustering correction, which we shall do in the next section, where we shall consider the full $C_F C_A$ result including that from the $C_F \left(C_F-C_A/2\right)$ term.

 \section{Structure of NLL results\label{sec:discussion}}

 In this section we  discuss the structure of our results for each of the $C_F^2$, $C_F C_A$ and $C_F T_R n_f$ channels. As has already been noted for every channel, the result at order $\alpha_s^2$ reproduces the terms expected from the expansion of the LL formula \eqref{eq:llexpansion} in addition to producing genuine NLL corrections.\footnote{Recall that the expansion of the LL formula also contains formally NLL terms but which can be embedded within the LL strongly ordered dynamics.} Therefore we may focus only on the additional terms not produced as part of the LL expansion and hence we write:
 \begin{equation}
 \label{eq:beyondll}
 \rho \frac{d\Sigma_2}{d\rho} = \rho \frac{d\Sigma^{\mathrm{LL}}_2}{d\rho} + \rho \frac{d\Sigma^{\mathrm{NLL}}_2}{d\rho},
 \end{equation}
 where $\rho \frac{d\Sigma^{\mathrm{LL}}_2}{d\rho}$ is reported in Eq.~\eqref{eq:llexpansion} and $ \rho \frac{d\Sigma^{\mathrm{NLL}}_2}{d\rho}$ describes the NLL terms unrelated to LL dynamics.

 To obtain our result for the $\rho \frac{d\Sigma^{\mathrm{NLL}}_2}{d\rho}$ in the $C_F^2$ channel we combine the result from gluon emission Eq.~\eqref{eq:emission} with the $C_F^2$ term arising from the gluon decay \eqref{eq:identical} identical particle piece. After removal of the LL contribution we can write:
 \begin{equation}
 \label{eq:cf2nll}
\begin{split}
 &\left(\rho \frac{d\Sigma^{\mathrm{NLL}}_2}{d\rho}\right)^{C_F^2} = \left(\frac{C_F  \alpha_s}{2\pi} \right)^2 \times\\
 &\left(2\pi^2  -24 \zeta(3) +\frac{1}{2}
 +\left(1.866 \pm 0.002 \right)  + \left ( 1.4386 \pm 0.0001     \right)  \right)  \\
 &+\mathcal{F}^{\mathrm{clust.}}_{C_F^2}\,,
\end{split}
 \end{equation}
 where $\mathcal{F}^{\mathrm{clust.}}_{C_F^2}$ is the clustering contribution Eq.~\eqref{eq:cluster}.

 The corresponding result for the $C_F  T_R n_f$ term can be obtained partly in numerical form from Eq.~\eqref{eq:nfres} or fully analytically from Eq.~\eqref{eq:nffinal2} and after removal of the LL contribution we get

 \begin{equation}
 \label{eq:cfnfnll}
 \left(\rho \frac{d\Sigma^{\mathrm{NLL}}_2}{d\rho}\right)^{C_F T_R n_f} = C_F T_R n_f \left(\frac{\alpha_s}{2\pi} \right)^2 \times 5.0+ \mathcal{F}^{\mathrm{clust.}}_{C_F  T_R n_f},
 \end{equation}
 where we used above the analytical result of Eq.~\eqref{eq:nffinal2} and the clustering contribution is given in Eq.~\eqref{eq:nfclust}.

 The result for the $C_F C_A$ channel is obtained by combining Eq.~\eqref{eq:CAfinal} with the identical particle contribution in Eq.~\eqref{eq:identical} and removing the LL contributions so that we obtain
 \begin{equation}
 \label{eq:CFCAnll}
 \begin{split}
 \left(\rho \frac{d\Sigma^{\mathrm{NLL}}_2}{d\rho}\right)^{C_F C_A}  &= C_F C_A  \left(\frac{\alpha_s}{2\pi} \right)^2
 \bigg(  16 \zeta (3)-\frac{11 \pi ^2}{9} \\
 &-\frac{121}{6}+(6.678\pm0.006) \\
 &- \left(0.7193  \pm  0.00005 \right)\bigg) +\mathcal{F}^{\mathrm{clust.}}_{C_F  C_A} \ \ ,
 \end{split}
 \end{equation}
 where $\mathcal{F}^{\mathrm{clust.}}_{C_F  C_A}$ is reported in \eqref{eq:CAclustfinal}.

 Leaving aside the clustering corrections for the moment, whose soft (and collinear) origin we have already discussed, we focus on the structure of the rest of the result. It is well-known that the intensity of collinear radiation from a quark at second order in $\alpha_s$ is related to a coefficient in the quark form factor generally referred to as $B^{(2)}$ \cite{Collins:1981uk,Collins:1981va,Kodaira:1981nh,Kodaira:1982az}. While there is not a unique definition of $B^{(2)}$ since it depends on the details of how the full resummation formula is organised, \ie the resummation scheme, it is always related to the endpoint $\delta(1-z)$ contribution to the NLO DGLAP splitting functions via the form (for a quark initiated jet) \cite{deFlorian:2000pr,Catani:2000vq,deFlorian:2004mp},
 \begin{equation}
 \label{eq:b2}
 B^{(2)} =  -2\gamma_q^{(2)}+C_F b_0 X,
 \end{equation}
 where $b_0=\frac{1}{6}\left(11 C_A -4 T_R n_f\right)$  and where $\gamma_q^{(2)}$, the DGLAP endpoint contributions for a quark jet, are \cite{Furmanski:1980cm,Curci:1980uw}
 \begin{equation}
 \label{eq:g2}
\begin{split}
\gamma_q^{(2)} &= C_F^2 \left(\frac{3}{8} -\frac{\pi^2}{2}+6 \zeta(3)  \right)\\
&+C_F C_A \left(\frac{17}{24}+\frac{11 \pi^2}{18} -3 \zeta(3) \right)
-C_F T_R n_f \left(\frac{1}{6}+\frac{2\pi^2}{9} \right).
\end{split}
\end{equation}
 We note that our analytic result for the coefficient of \\ $C_F T_R n_f \alpha_s^2/(2\pi)^2$ in Eq.~\eqref{eq:cfnfnll} is precisely consistent with the form in Eq.~\eqref{eq:b2} with
 \begin{equation}
 \label{eq:x}
 X= \frac{2\pi^2}{3}-7.
 \end{equation}
 Taking this value of $X$ we obtain a result $-7.03766$  for the  $C_F C_A$ term in Eq.~\eqref{eq:b2} in good agreement with the numerical value for our result in Eq.~\eqref{eq:CFCAnll}  without the clustering corrections, where we get  $-7.03791 \pm 0.006$. Finally Eq.~\eqref{eq:b2} gives a value $-5.30508$ for the coefficient of $C_F^2$, which again agrees well with the result in  Eq.~\eqref{eq:cf2nll} without the clustering correction, which has the numerical value \\ $-5.30556  \pm 0.002$.  Furthermore, using the de Florian and Grazzini analytical result (Eq.~\eqref{eq:idanalyt}), for the identical particle gluon decay contribution, we can identify the remaining numerical contributions with analytic results i.e. the result $1.866$ for the $C_F^2$ piece in Eq.~\eqref{eq:cf2nll} corresponds to $8 \zeta(3)-\frac{31}{4}$ while the result
 $6.678$ in Eq.~\eqref{eq:CFCAnll} corresponds to \\ $\frac{13  \pi^2}{18} -8 \zeta(3)+\frac{55}{6}$.

 We conclude that without the clustering corrections our NLL results for the mMDT as defined above, are given by  the general collinear form Eq.~\eqref{eq:b2} with the  value of $X$ specified in \eqref{eq:x}. Thus excluding the clustering corrections our NLL result for the mMDT has a simple correspondence to the collinear order $\alpha_s^2$ contribution to the quark form factor. Moreover the results for the $C_F C_A$  and $C_F  T_R n_f$ channels, without the clustering correction, agree exactly with the order $\alpha_s^2$ NNLL term in the expansion of the ungroomed heavy jet mass \cite{Chien:2010kc,Banfi:2014sua,Banfi:2018mcq}, a consequence of grooming affecting only soft emissions in the small $\zc$ limit.  Finally we note that our overall results in every channel are in agreement with those from previous SCET calculations. An explicit expansion of the SCET results to order $\alpha_s^2$, for the jet mass distribution, has recently been provided in Ref.~\cite{Kardos:2020ppl}.\footnote{Note that we have removed $\ln 2$ terms in the $C_F C_A$ and $C_F n_f$ channels present in the results of \cite{Kardos:2020ppl} via the choice of $Q^2/4$ in the scale of $\alpha_s$ for our leading-order result.}

 \section{Conclusions}
 In this article we have revisited the NLL structure of the jet mass distribution for mMDT groomed jets from the viewpoint of its direct  connection to the QCD matrix elements in the triple-collinear limit. Previous NLL results have entirely been within the framework of Soft-Collinear Effective Theory (SCET) and hence our work represents an approach which provides strong independent confirmation of the main results involved in the NLL resummation \cite{Frye:2016aiz,Bell:2018vaa,Chien_2016,von_Manteuffel_2014}.

Our results establish a connection between the NLL \\ groomed jet mass result and the standard ingredients used in QCD resummation. In particular we recover the expected scale of the running coupling in the soft limit, i.e. the $k_t$ of a soft emission, and the constant $K$ related to the emergence of the CMW coupling. We further obtain a link between the NLL result and the general form of the $B^{(2)}$ coefficient that controls the intensity of collinear radiation from a quark at order $\alpha_s^2$ and hence enters the quark form factor. Our result also involves a clustering correction in all channels, which stems from our simplifying the action of the mMDT to derive the $B^{(2)}$-like terms. The clustering corrections come from a region of phase space where we have two soft emissions that if examined individually fail the $\zc$ condition but if examined as a cluster pass the $\zc$ condition. We believe that these results should allow for a resummation of the mMDT NLL corrections within a QCD resummation framework. In fact  the $B^{(2)}$-like pieces are already incorporated in an approach such as ARES \cite{Banfi:2014sua,Banfi:2018mcq} as they also enter into the NNLL structure of the ungroomed heavy jet mass.\footnote{In ARES \cite{Banfi:2014sua,Banfi:2018mcq}, for the heavy jet mass the factor $X$ in the definition used for $B^{(2)}$ is equal to zero and the $b_0 X$ term is associated to the functions $C^{(1)}_{\text{hc}}$ and $\delta\mathcal{F}_{\text{rec}}$.} It therefore remains to consistently include the clustering corrections within the standard QCD resummation formalism.

 For future extensions of our work, one development that is possible to make concerns the inclusion of finite $\zc$ corrections beyond the LL level \cite{Dasgupta:2013ihk,Marzani:2017mva, Marzani:2017kqd}. These can be derived through our triple-collinear calculations retaining terms that we have omitted in the present article by taking the small $\zc$ limit of various formulae. Since these additional corrections will be purely finite, they can be computed numerically in four dimensions and incorporated into the resummation framework described in Ref.~\cite{Dasgupta:2013ihk}. While we expect the resulting corrections to be numerically small, they should become relevant to examine in the context of recent developments pushing the mMDT jet mass resummation to NNLL level \cite{Kardos:2020gty}. It would also be of interest to use  our approach to study $\beta \neq 0$ values for SoftDrop and hence to develop an NNLL QCD resummation approach for those observables.  Our triple-collinear calculations should also give the insight needed to address other similar collinear problems at the NLL level such as that involving the small jet radius limit of QCD jets, for which an LL resummation formalism was constructed in Ref. \cite{Dasgupta:2014yra} but a general resummation approach at NLL is still missing.

\begin{acknowledgements}
We would like to thank Andrea Banfi, Fr\'ed\'eric Dreyer, Keith Hamilton, Pier Monni, Gavin Salam, Gregory Soyez and Mike Seymour for useful discussions. This work has been funded by the European Research Council (ERC) under the  European Unions Horizon 2020 research and innovation programme (grant agreement No. 788223, PanScales) (MD, BKE) and by the Science and Technologies Facilities Council (STFC) under grant ST/P000800/1 (MD).
The work of MG is supported by the National Science Foundation under Grant No. PHY 1820818.
MD would like to thank CERN for a scientific associateship award during the course of which this work was initiated.
JH thanks the UK Science and Technologies Facilities Council (STFC) for a PhD studentship award.
\end{acknowledgements}

\appendix

\label{sec:appendix}
\section{Triple-Collinear splitting functions and integrals}
\label{sec:tripcoll}
We use the results in the form listed in Refs.~\cite{Catani:1998nv,Catani:1999ss}. Following the notation of those references we define $\mathcal{T}$
as the squared matrix element for $e^+e^-\rightarrow 4$ partons. In the limit where three of the final partons are collinear, it can be shown that $\mathcal{T}$ satisfies the following factorized form:
\begin{equation}
\begin{split}
\label{eq:tripcollfac}
&\mathcal{T}(e^+e^-\rightarrow 4~\mathrm{partons})\simeq\mathcal{T}(e^+e^-\rightarrow q\bar q)\cdot\sum_k\mathcal{T}_k^\mathrm{coll}(1\rightarrow3)
\\
&=\mathcal{T}(e^+e^-\rightarrow q\bar q)\cdot\frac{(8\pi\alpha_s\mu^{2\epsilon})^2}{s_{123}^2}\sum_k\la\hat P^k_{1\rightarrow3}\ra.
\end{split}
\end{equation}
Here, the $k$ runs over the possible quark initiated $1\rightarrow3$ parton channels $q\rightarrow g_1 g_2 q_3~(\bar q\rightarrow g_1 g_2 \bar q_3),~q\rightarrow q'_1 \bar q'_2 q_3~(\bar q\rightarrow \bar q'_1 q' _2\bar q_3)$ or $q\rightarrow q_1 \bar q_2 q_3~(\bar q\rightarrow \bar q_1 q_2 \bar q_3)$, $s_{123}$
is the squared invariant mass of the three collinear parton system, and $\la\hat P^k_{1\rightarrow3}\ra$ are the process independent spin averaged triple-collinear splitting functions.

Hereinafter, we report the relevant expressions for the triple-collinear splitting functions~\cite{Catani:1998nv}. Note that due to charge conjugation invariance, the splitting functions for the anti-quark initiated channels can be obtained from the corresponding functions for the quark initiated ones,. i.e.
$\Ph_{{\bar q}^\prime_1 q^\prime_2 {\bar q}_3} =
\Ph_{{\bar q}^\prime_1 q^\prime_2 q_3}$ and
$\Ph_{{\bar q}_1 q_2 {\bar q}_3} =
\Ph_{{\bar q}_1 q_2 q_3}$. \\

The spin-averaged splitting function for the $q\rightarrow g_1 g_2 q_3$ process can be
written in terms of the different colour factors:
\beq
\label{qggsf}
\la \Ph_{g_1 g_2 q_3} \ra \, =
C_F^2 \, \la \Ph_{g_1 g_2 q_3}^{({\rm ab})} \ra \,
+ \, C_F C_A \, \la \Ph_{g_1 g_2 q_3}^{({\rm nab})} \ra  \;\;,
\eeq
where the abelian and non-abelian contributions are
\beeq
\label{eq:qggabsf}
\la \Ph_{g_1 g_2 q_3}^{({\rm ab})} \ra \,
&=&\Biggl\{\f{s_{123}^2}{2s_{13}s_{23}}
z_3\left[\f{1+z_3^2}{z_1z_2}-\ep\f{z_1^2+z_2^2}{z_1z_2}-\ep(1+\ep)\right]\nn\\
&+&\f{s_{123}}{s_{13}}\Biggl[\f{z_3(1-z_1)+(1-z_2)^3}{z_1z_2}+\ep^2(1+z_3)
\nn\\
&-&\ep (z_1^2+z_1z_2+z_2^2)\f{1-z_2}{z_1z_2}\Biggr]\nn\\
&+&(1-\ep)\left[\ep-(1-\ep)\f{s_{23}}{s_{13}}\right]
\Biggr\}+(1\lra 2) \;\;,
\eeeq

\beeq
\label{qggnabsf}
&&\la \Ph_{g_1 g_2 q_3}^{({\rm nab})} \ra \,
=\Biggl\{(1-\ep)\left(\f{t_{12,3}^2}{4s_{12}^2}+\f{1}{4}
-\f{\ep}{2}\right)\nn\\
&+&\f{s_{123}^2}{2s_{12}s_{13}}
\Biggl[\f{(1-z_3)^2(1-\ep)+2z_3}{z_2}
+\f{z_2^2(1-\ep)+2(1-z_2)}{1-z_3}\Biggr]\nn\\
&-&\f{s_{123}^2}{4s_{13}s_{23}}z_3\Biggl[\f{(1-z_3)^2(1-\ep)+2z_3}{z_1z_2}
+\ep(1-\ep)\Biggr]\nn\\
&+&\f{s_{123}}{2s_{12}}\Biggl[(1-\ep)
\f{z_1(2-2z_1+z_1^2) - z_2(6 -6 z_2+ z_2^2)}{z_2(1-z_3)}\nn\\
&+&2\ep\f{z_3(z_1-2z_2)-z_2}{z_2(1-z_3)}\Biggr]\nn\\
&+&\f{s_{123}}{2s_{13}}\Biggl[(1-\ep)\f{(1-z_2)^3
+z_3^2-z_2}{z_2(1-z_3)}\nn\\
&-&\ep\left(\f{2(1-z_2)(z_2-z_3)}{z_2(1-z_3)}-z_1 + z_2\right)\nn\\
&-&\f{z_3(1-z_1)+(1-z_2)^3}{z_1z_2}
+\ep(1-z_2)\left(\f{z_1^2+z_2^2}{z_1z_2}-\ep\right)\Biggr]\Biggr\}\nn\\
&+&(1\lra 2) \;\;.
\eeeq
The spin-averaged splitting functions for non-identical fermions in the final state read
\begin{equation}
\label{qqqpf}
\begin{split}
\la \Ph_{{\bar q}^\prime_1 q^\prime_2 q_3} \ra \, = \f{1}{2} \,
C_F T_R \,\f{s_{123}}{s_{12}} &\left[ - \f{t_{12,3}^2}{s_{12}s_{123}}
+\f{4z_3+(z_1-z_2)^2}{z_1+z_2}\right.\\
&\left.+ (1-2\ep) \left(z_1+z_2-\f{s_{12}}{s_{123}}\right)
\right] \;,
\end{split}
\end{equation}
where
\begin{equation}\label{tvar}
t_{ij,k}\equiv 2 \,\frac{z_i s_{jk}-z_j s_{ik}}{z_i+z_j} +
\frac{z_i-z_j}{z_i+z_j} \,s_{ij} \;\;.
\end{equation}
In the case of final-state fermions with
identical flavour, the splitting function can be written in terms of Eq. (\ref{qqqpf}),
as
\beq\label{qqqsf}
\la \Ph_{{\bar q}_1q_2q_3} \ra \, =
\left[ \la \Ph_{{\bar q}^\prime_1q^\prime_2q_3} \ra \, + \,(2\lra 3) \,\right]
+ \la \Ph^{({\rm id})}_{{\bar q}_1q_2q_3} \ra \;\;,
\eeq
where
\begin{equation}\label{idensf}
\begin{split}
\la \Ph^{({\rm id})}_{{\bar q}_1q_2q_3} \ra \,
&= C_F \left( C_F-\f{1}{2} C_A \right)
 \Biggl\{ (1-\ep)\left( \f{2s_{23}}{s_{12}} - \ep \right)\\
&+ \f{s_{123}}{s_{12}}\Biggl[\f{1+z_1^2}{1-z_2}-\f{2z_2}{1-z_3}\\
&-\ep\left(\f{(1-z_3)^2}{1-z_2}+1+z_1-\f{2z_2}{1-z_3}\right)
- \ep^2(1-z_3)\Biggr] \\
&- \f{s_{123}^2}{s_{12}s_{13}}\f{z_1}{2}\left[\f{1+z_1^2}{(1-z_2)(1-z_3)} \right.\\
&\left.-\ep \left(1+2\f{1-z_2}{1-z_3}\right)
    -\ep^2\right] \Biggr\} + (2\lra 3)\,.
\end{split}
\end{equation}

\section{Integrals and pole structure \label{sec:integrals}}

We illustrate our approach by referring explicitly to the calculations for $\mathcal{F}^{\text{pass}}(\rho,\epsilon)$, though similar considerations apply to all our calculations. When all three partons contribute to the jet mass i.e. we have the delta function condition $\delta_\rho(1,2,3)$, we can eliminate the integral over an angle, say $\theta_{12}$, by  using the delta function condition. In order to perform the remaining integrals over the triple-collinear phase-space $d\Phi_3$, (see  Eq.~\eqref{eq:tcphasesp}) it proves to be convenient to exploit the fact that all partons are collinear, with a collinearity essentially set by the jet mass or more accurately by the parameter $\frac{\rho}{\zc}$. One may extract the overall $1/\rho$ scaling of $d\Sigma/d\rho$  by working in terms of rescaled angular variables $y=  \frac{{\theta^2_{23}}}{{\theta^2_{13}}}$ and  $x =\frac{\theta_{13}^2}{\rho}$. The limits on the $x$ integral follow from the positivity of the Gram Determinant $\Delta>0$, corresponding to the conditions $\left (\theta_{13}+\theta_{23}  \right)^2  > \theta^2_{12} > \left ( \theta_{13}-\theta_{23}  \right)^2$, which can be expressed in terms of our chosen variables as a condition on $x$:
\begin{equation}
\begin{split}
&\frac{1}{z_1(1-z_1)+y z_2 (1-z_2) +2 \sqrt{y} z_1 z_2} < x \,\, , \\
&x <  \frac{1}{z_1(1-z_1)+y z_2 (1-z_2) -2 \sqrt{y} z_1 z_2}\,\,.
\end{split}
\end{equation}
It proves to be convenient to map the integral over $x$ to one with simple limits i.e. $0$ and $1$ by introducing the change of variables $x=u(r_2-r_1)+r_1$. Our integration variables are then  $u$ and $y$ for the angular integration, both lying in a range 0 to 1, and the energy fractions $z_1$ and $z_2$ (recall that  $z_3=1-z_1-z_2$) or equivalently $z$ and $z_p$.

We then have to  consider the extraction of $\epsilon$ poles, to separate the integral into divergent and finite terms. Our strategy is to isolate the divergences and exploit the simplification of the integrand in divergent  regions, to obtain  the divergence structure analytically. This also generates finite terms that do not vanish as $\epsilon \to 0$ , which are obtained via an  $\epsilon$ expansion of the factors multiplying the poles.  Additionally we also obtain a finite integral leftover from the removal of singular terms, which on the other hand is not a compact expression. However, being finite, it can always be integrated numerically.

Since we study the differential distribution rather than its integral, we work at fixed jet-mass which regulates both soft and collinear divergence. In general that leaves us with at most a $1/\epsilon^2$ singularity from an emission that does not set the jet mass. For the calculation of $\mathcal{F}^{\text{pass}}$  in particular, the larger-angle emission passes $\zc$ and cannot produce any divergence,  while the smaller angle emission produces divergences from the soft $z_p \to 1$ and collinear $y \to 0$ limits. Setting $1-z_p =v$ we encounter a general integral of the standard form
\begin{equation}
I(\epsilon) = \int_0^1 dv \int_0^1 dy \frac{G \left(v,y,\epsilon \right)}{v^{1+2\epsilon} y^{1+\epsilon}},
\end{equation}
where $G(v,y,\epsilon)$ is finite as $v \to 0$ as well as $y \to 0$ and integration over the other variables is left implicit so as to focus on the divergences.
We can re-express this result in the following form
\begin{equation}
\begin{split}
I(\epsilon) =  \int_0^1 dv \int_0^1 dy  &\left [ \frac{G(v,y,\epsilon)-G(v,0,\epsilon)}{v^{{1+2\epsilon}} y^{1+\epsilon}}
\right.\\
&\left.+ \frac{G(v,0,\epsilon)-G(0,0,\epsilon)}{v^{{1+2\epsilon}} y^{1+\epsilon}}  +\frac{G(0,0,\epsilon)} {v^{{1+2\epsilon}} y^{1+\epsilon}}\right ],
\end{split}
\end{equation}
where by construction the first term on the LHS of the above has only a soft pole i.e. as $v \to 0$, the second term has only a collinear pole from $y \to 0$, while the final term has a double pole arising from $v \to 0$ and $y \to 0$.  We define for convenience $f(v,\epsilon) = \left(G(v,0,\epsilon)-G(0,0,\epsilon) \right)/v^{1+2\epsilon}$ where $f(v,\epsilon)$ is finite as $v \to 0$ and also define
\[h(v,y,\epsilon) = \left(G(v,y,\epsilon)-G(v,0,\epsilon) \right)/y^{1+\epsilon},\]
which is finite as $y \to 0$.  Then one obtains the form
\begin{equation}
\begin{split}
I(\epsilon) &= \frac{G(0,0,\epsilon)}{2\epsilon^2} -\frac{1}{\epsilon}  \int_0^1 f(v,\epsilon) dv
-\frac{1}{2\epsilon} \int_0^1 h(0,y,\epsilon) dy \\
&+ \int_0^1 dv \int_0^1 \frac{h(v,y,\epsilon)-h(0,y,\epsilon)}{v^{1+2\epsilon}} dy \ \ .
\end{split}
\end{equation}
The final integral on the RHS above is purely finite by construction and can be evaluated in the limit $\epsilon \to 0$ i.e in 4 dimensions.  The above result shows explicitly the pole structure that emerges from the integral $I(\epsilon)$. The integrals multiplying the $1/\epsilon$ poles need only to be evaluated up to order $\epsilon$ terms i.e. one can expand the integrand in $\epsilon$ and retain only terms up to order $\epsilon$. This strategy gives us all divergent and finite contributions in the limit $\epsilon \to 0$.

\section{The web variables}
\label{app:web}
We start with the full four-body phase space
\begin{equation}
\begin{split}
\text{d} \Phi_{(4)} &= \frac{\text{d}^{d-1} p}{(2\pi)^{d-1}} \frac{1}{ 2E_p}   \frac{\text{d}^{d-1} \bar{p}}{(2\pi)^{d-1}}  \frac{1}{ 2E_{\bar{p}}}
\times\\
&\frac{\text{d}^{d-1} k_1}{(2\pi)^{d-1}} \frac{1}{ 2 E_1 }   \frac{\text{d}^{d-1} k_2}{(2\pi)^{d-1}} \frac{1}{ 2 E_2 }\, \frac{1}{ 2E_{\bar{p}}}\,
\\
&\times (2\pi)^d \delta^{(d)} (Q- p -\bar{p}- k_1 - k_2) \,,
\end{split}
\end{equation}
where the momentum of the quark is denoted by $p$, the anti-quark by $\bar{p}$ and the emitted partons by $k_1$ and $k_2$. Without loss of generality, we take the partons to be collinear to the quark direction and thus integrate over the anti-quark three-momentum to collapse the spatial delta function. We find
\begin{equation}
\begin{split}
\text{d} \Phi_{(4)} &= \frac{\text{d}^{d-1} p}{(2\pi)^{d-1}} \frac{1}{ 2E_p}
  \frac{\text{d}^{d-1} k_1}{(2\pi)^{d-1}} \frac{1}{ 2 E_1 }\times
\\
  &\frac{\text{d}^{d-1} k_2}{(2\pi)^{d-1}} \frac{1}{ 2 E_2 }\, \frac{1}{ 2E_{\bar{p}}}\, (2\pi) \delta (Q - E_p - E_{ \bar{p}} - E_1 -E_2) \ \ ,
\end{split}
\end{equation}
where
\begin{align}
E_{\bar{p}} = \sqrt{E_p^2 + E_1^2 +E_2^2 + 2 \vec{k}_1 \cdot \vec{p} + 2 \vec{k}_2 \cdot \vec{p}+ 2 \vec{k}_1 \cdot \vec{k}_2} \ \ .
\end{align}
Now in the triple-collinear limit all pairwise angles are small, which allows the following simplification
\begin{equation}\label{eq:tcbasic}
\begin{split}
\text{d} \Phi^{\text{tc}}_{(4)} &= (2\pi)^{2-d} \frac{d\Omega_{d-1}}{4Q}  \frac{\text{d}^{d-1} k_1}{(2\pi)^{d-1}} \frac{1}{ 2 E_1 }   \frac{\text{d}^{d-1} k_2}{(2\pi)^{d-1}} \frac{1}{ 2 E_2 }\times \\
&E_p^{1-2\epsilon} dE_p \, \delta \left(\frac{Q}{2} - E_p - E_1 -E_2 \right)  \ \ ,
\end{split}
\end{equation}
where the delta function forces the expected constraint on the energies of the three-parton system. Notice that  without this energy constraint the triple-collinear phase space looks identical to an {\em unconstrained} 2-body phase space.

Now we come to the most important step. We want to re-express $\text{d} \Phi^{\text{tc}}_{(4)}$ in terms of the transverse momentum of the parent gluon, with respect to the quark direction, in addition to the invariant mass of the composite parton system, i.e. $k_1 + k_2$. To this aim we introduce the Sudakov decomposition along the quark direction and the associated phase space for a massless parton
\begin{equation}
\begin{split}
&k = x p + \bar{x} \bar{p}_s + k_\perp ,\quad k_t^2 = - k_\perp^2, \quad
\\
&\frac{\text{d}^{d-1} k}{(2\pi)^{d-1}} \frac{1}{ 2 E } = \frac{1}{(2\pi)^{d-1}} \frac{dx}{2x} d^{d-2}k_\perp \ \ .
\end{split}
\end{equation}
where $\bar{p}_s$ is a spectator momentum given by $\bar{p}_s=(E_p,-\vec{p})$. Here, $k_\perp $ is the transverse momentum with respect to the quark direction. In the collinear approximation the relation between the Sudakov $x$ and the energy of the emissions is quite simple, $x_i = E_i/E_p$ , and thus eq.~\eqref{eq:tcbasic} becomes
 \begin{equation}\label{eq:tcsud}
\begin{split}
 \text{d} \Phi^{\text{tc}}_{(4)} &= \frac{ (2\pi)^{4-3d}}{16Q} d\Omega_{d-1} \frac{dE_1}{ E_1 } \frac{dE_2}{ E_2 } d^{d-2}k_{\perp1}d^{d-2}k_{\perp2} \,
\\
& \times E_p^{1-2\epsilon} dE_p \, \delta \left(\frac{Q}{2} - E_p - E_1 -E_2 \right)  \ \ .
 \end{split}
 \end{equation}
We change to energy fractions, viz. $ E_1 = Q/2 (1-z) z_p$ and $E_2 = Q/2 (1-z) (1-zp)$ and
use the delta function to integrate over $E_p$ to obtain
\begin{equation}
\begin{split}
  \text{d} \Phi^{\text{tc}}_{(4)} &= \frac{ (2\pi)^{4-3d}}{16Q} d\Omega_{d-1} \left(\frac{Q}{2}\right)^{1-2\epsilon}
\\
&  \times\frac{ z^{1-2\epsilon} dz}{1-z} \frac{dz_p}{z_p(1-z_p)}\, d^{d-2}k_{\perp1}d^{d-2}k_{\perp2}  \ \ .
\end{split}
\end{equation}
Finally, we introduce the transverse momentum of the parent gluon and the invariant mass of the emissions as follows
\begin{equation}
\begin{split}
 &k_\perp = k_{\perp 1} + k_{\perp 2}, \quad q_\perp = \frac{k_{\perp 1}}{z_p} - \frac{k_{\perp 2}}{1-z_p}, \quad
 \\
& z_p(1-z_p) q_t^2 = s_{12} \ \ .
 \end{split}
 \end{equation}
Implementing the above transformations eq.~\eqref{eq:tcsud} takes on the relatively simple form
 \begin{align}
   \text{d} \Phi^{\text{tc}}_{(4)} =    \text{d} \Phi_{\text{B}} \times \text{d} \Phi_3 \ \ ,
   \end{align}
where $ \text{d} \Phi_{\text{B}}$ is the two-body phase space that we must extract to form the Born cross section, viz.
\begin{align}\label{eq:born}
  \text{d} \Phi_{\text{B}} = \frac{4^\epsilon (2\pi)^{2-d}}{8 Q^{2\epsilon}} d\Omega_{d-1} \ \ ,
 \end{align}
while the triple-collinear phase space reads
 \begin{equation}\label{eq:tcfinal}
 \begin{split}
 \text{d} \Phi_3 &= \frac{(4\pi)^{2\epsilon}}{256 \pi^4} \frac{2 z^{1-2\epsilon} dz}{1-z} \frac{1}{\Gamma(1-\epsilon)} \frac{d^{2-2\epsilon}k_\perp}{\Omega_{2-2\epsilon}} \frac{ds_{12}}{(s_{12})^\epsilon}
\\
&\times \frac{dz_p}{(z_p(1-z_p))^\epsilon} \frac{1}{\Gamma(1-\epsilon)} \frac{d\Omega_{2-2\epsilon}}{\Omega_{2-2\epsilon}}  \ \ .
\end{split}
\end{equation}
Here, the solid angle $d\Omega_{2-2\epsilon}$ is that of $q_\perp$ in the transverse plane aligning $k_\perp$ along one axis. A nice feature of eq.~\eqref{eq:tcfinal} is the simplicity of the double-soft limit, $z \to 1$. In fact, if we set $z=1$ in the measure one recovers the double-soft phase space, the 4-dimensional limit of which is reported in \cite{Dokshitzer:1998pt}.

\section{One-loop corrections to $1\to 2$ collinear splittings \label{sec:r1v1}}

In addition to the case of two real emissions, for the jet mass distribution at order $\alpha_s^2$ we also have to consider a real emission that sets the mass $\rho$ alongside a one-loop virtual correction which is divergent and where the divergences cancel against those in the double-real emission case, to leave behind finite terms. The relevant real-virtual contribution to $\rho d\Sigma_2/d\rho$, may be collectively written in the form
\begin{equation}\label{eq:VirtualsCF}
\begin{split}
\mathcal{V}_{1,1} \left(\rho,z,\epsilon \right) &= \mathcal{V}_{1,1}^{C_F^2}(\rho, z,\epsilon)+\mathcal{V}_{1,1}^{C_FC_A}(\rho,z,\epsilon)
\\
&+ \mathcal{V}_{1,1}^{C_FT_R n_f}\left(\rho,z,\epsilon\right),
\end{split}
\end{equation}
where we have separated out the various contributions according to the colour factor i.e. $C_F^2$, $C_F C_A$ and $C_F  T_R n_f$ terms and in our notation $\mathcal{V}_{1,1}$ is the one-real, one-virtual correction to $q \bar{q}$ production, in the approximation of a real emission, which is collinear to the $q$ or $\bar{q}$, passes grooming and sets a (normalised) jet mass $\rho$.

For  $\mathcal{V}_{1,1}^{C_F^2}(\rho, z,\epsilon)$ there are two distinct contributions : firstly the one-loop correction to the Born level $q\bar{q}$ production, $\mathcal{V}(\epsilon)$ (see Eq.~\eqref{eq:virtborn}) multiplying the squared matrix-element for a real collinear emission, and secondly the one-loop correction to a $1 \to  2$ collinear splitting \cite{Sborlini:2013jba}. The latter contribution can be explicitly obtained using the expression for
$P_{q\to qg}^{(1)}$ in Eq.~(103) of Ref.~\cite{Sborlini:2013jba}, in the CDR  scheme with $\alpha  =1$ and  $\delta=1$, and setting $s_{12} = \frac{Q^2}{4} \rho$, $\mu^2=Q^2/4$ and $z_1=1-z$.  The Hypergeometric function in Eq.~(103) of  Ref.~\cite{Sborlini:2013jba} may be expressed in terms of a function
\[f\left(\epsilon,1/x\right)  =  \frac{1}{\epsilon}\left[ _2 F_1\left(1,-\epsilon,1-\epsilon,1-x\right)-1\right], \]
with the  $\epsilon$ expansion \cite{Catani:2011st}
\begin{equation}
f(\epsilon,x) = \ln x-\epsilon \left [\text{Li}_2(1-x)+\sum_{k=1}^{\infty}  \epsilon^k  \, \text{Li}_{k+2}(1-x)\right].
\end{equation}
Writing our result in terms of the renormalised $\overline{\mathrm{MS}}$  coupling, accounting for both hemispheres with a factor of two, we have
\begin{multline}
\label{eq:r1v1}
\mathcal{V}_{1,1}^{C_F^2}(\rho,z,\epsilon) =  \left(\frac{C_F \alpha_s}{2\pi}\right)^2  \left [
\vphantom{\mathrm{Li}_2\left(\frac{z-1}{z}\right)}
2  \, p_{qq}(z;\epsilon)\rho^{-\epsilon}(4z(1-z))^{-\epsilon}
\right.\\
\left.\times\left( -\frac{2}{\epsilon^2} +\frac{4\pi^2}{3}-8 -\frac{3}{\epsilon} \right)+ p_{qq}(z,\epsilon) \rho^{-2 \epsilon}(z(1-z))^{-\epsilon}
\right. \\
\left.
\times \left(\frac{4}{\epsilon}\ln z+4\ \mathrm{Li}_2\left(\frac{z-1}{z}\right)\right)-2 \right].\\
\end{multline}
The corresponding result for the $C_F T_R n_f$ piece, after removal of UV poles via renormalisation (see e.g. Ref.~\cite{Catani:2011st} for a detailed discussion), can be expressed as
\begin{equation}
 \label{eq:nfvirtual}
 \begin{split}
 \mathcal{V}_{1,1}^{C_FT_R n_f}\left(\rho,z,\epsilon\right) &= 2 C_F T_R n_f  \left(\frac{\alpha_s}{2\pi}\right)^2 \frac{2}{3} \times
\\
& p_{qq}(z,\epsilon)\left(\frac{1}{\epsilon}\rho^{-2\epsilon}(z(1-z))^{-\epsilon}+\ln\rho\right).
\end{split}
\end{equation}

The result for the  $C_F C_A$ piece has two distinct components i.e. a component derived like the corresponding $C_F^2$ piece using Eq.~(103) of  Ref.~\cite{Sborlini:2013jba} which includes the double poles that will cancel those in the real emission result, and a component involving the $\beta$ function coefficient $b_0$ which is simply related to Eq.~\eqref{eq:nfvirtual} via the replacement $\frac{2}{3} T_R n_f \to  - \frac{11}{6} C_A$. The combined result can be expressed in the form
\begin{multline}
 \label{eq:cavirtual}
 \mathcal{V}_{1,1}^{C_F C_A}\left(\rho,z,\epsilon\right) = 2 C_F C_A  \left(\frac{\alpha_s}{2\pi}\right)^2\left\{
\vphantom{\mathrm{Li}_2\left(\frac{z-1}{z}\right)}
 p_{qq}(z,\epsilon) \rho^{-2\epsilon}  (z(1-z))^{-\epsilon}
\right.\\
\left. \times\left[\frac{-1}{\epsilon^2}+\frac{1}{\epsilon}\left(\ln \frac{1-z}{z}-\frac{11}{6}\right) \right] +  p_{qq}(z)  \left[ \mathrm{Li}_2 \left(\frac{z}{z-1} \right) -
\right.\right. \\  \left.\left.
\mathrm{Li}_2 \left(\frac{z-1}{z} \right) +\frac{2\pi^2}{3} -\frac{11}{6}\ln\rho\right]  +1 \right \}.
 \end{multline}

\bibliographystyle{spphys}
%\bibliography{bibliography}

\end{document}